\documentclass[twocolumn,aps,pra,a4paper,showpacs,preprintnumbers,amsmath,amssymb]{revtex4}

\usepackage[english]{babel}
\usepackage[dvips]{graphicx}
\usepackage{dcolumn}
\usepackage{bm}
\usepackage{amssymb}
\usepackage{mathrsfs}
\begin{document}
\title{Tomographic reconstruction of quantum correlations in excited Bose-Einstein condensates}

\author{Anders S. Mouritzen\footnote{Comments should be directed to this person}}
 \email{asm@phys.au.dk}
\author{Klaus M\o lmer}
 \email{moelmer@phys.au.dk}
\affiliation{QUANTOP, Danish National Research Foundation Center for
Quantum Optics, Department of Physics and Astronomy, University of
Aarhus, DK-8000 \AA rhus C, Denmark}

\date{\today}

\begin{abstract}
We propose to use quantum tomography to characterize the state of a
perturbed Bose-Einstein condensate. We assume knowledge of the
number of particles in the zero-wave number mode and of density
distributions in space at different times, and we treat the
condensate in the Bogoliubov approximation. For states that can be
treated with the Gross-Pitaevskii equation, we find that the
reconstructed density operator gives excellent predictions of the
second moments of the atomic creation- and annihilation operators,
including the one-body density matrix. Additional inclusion of the
momentum distribution at one point of time enables somewhat reliable
predictions to be made for the second moments for mixed states,
making it possible to distinguish between coherent and thermal
perturbations of the condensate. Finally, we find that with
observation of the zero-wave number mode's anomalous second moment,
$\langle \hat a_0 \hat a_0 + \hat a_0^\dagger \hat a_0^\dagger
\rangle$, the reconstructed density operator gives reliable
predictions of the second moments of locally amplitude squeezed
states.
\end{abstract}


\maketitle

\section{\label{intro}Introduction}
Bose-Einstein condensates offer countless demonstrations of
macroscopic quantum effects \cite{Cornell}, \cite{Ketterlenob}.
Recently, there has been growing interest in the use of condensates
as quantum optical components and as atom laser beam sources. The
effective use of very many of these applications requires knowledge
of certain aspects of the condensate's quantum state, and several
articles have dealt with theoretical calculations of the states
obtained by various perturbations of a condensate.

We present here a method by which experimental data can be used to
estimate characteristics of the system's quantum state. For this
purpose we generalize the application of the Jaynes principle of
maximum entropy (MAXENT) \cite{jaynesto} for quantum state
tomography, as proposed in \cite{Buzekjmo} and applied to single
atoms in \cite{Buzekatom} and \cite{Juhl}, to the dynamics of a
condensate described in the Bogoliubov approximation. We assume that
the fraction of particles in the zero-wave number mode and the
position distributions at different times after preparation of the
system have been measured. It is important to note that we are not
claiming that the MAXENT density operator is an approximation to the
true many-body operator of the quantum state; indeed this would
surely require more than just density measurements (one-body
operators). Instead, we seek to correctly predict the second moments
of the atomic creation- and annihilation operators $\hat \psi(x)$
and $\hat \psi^\dagger(x)$. Because these second moments of the
ladder operators will be discussed extensively below, we shall call
second moments of the form $\langle
\hat\psi^\dagger(x)\hat\psi^\dagger(x') \rangle$ and $\langle
\hat\psi(x)\hat\psi(x') \rangle$ anomalous second moments, while
second moments of the form $\langle \hat\psi^\dagger(x)\hat\psi(x')
\rangle$ will be referred to as normal second moments.

As an example of applying the method, three types of quantum
mechanically very different states will be studied. First, we study
states characterized by a single Gross-Pitaevskii wave function with
small deviations from the constant value attained by the ground
state homogeneous condensate. Then secondly, we apply our method to
an incoherently exited system, modeled by the addition of a
localized thermal component to the homogeneous condensate. Finally,
as a third case, we study the method's fidelity for a system with a
localized amplitude squeezed atomic field.

Our paper is organized as follows. In section \ref{teori} we give a
brief introduction to the central ideas behind tomography and a
description of the procedure for reconstructing the density operator
of a quantum system by use of Jaynes principle of maximum entropy.
In section \ref{sek:thecondensate} we specify our assumptions about
the condensate, and we introduce the Bogoliubov approximation and
the general para-unitary transformation procedure. In section
\ref{sek:minteori} we will combine the MAXENT and Bogoliubov
theories. In section \ref{sek:eksemplerne}, we apply the machinery
to the three different trial states. In section \ref{sek:disk} we
discuss the applicability of the method and we finally conclude the
paper in section \ref{sek:konklusion}.

\section{\label{teori}Quantum state tomography}
\subsection{Tomography\label{sek:tomografi}}
Can you find all the peaks of a mountain landscape or draw a map of
a city just by seeing its skylines from all compass directions? Of
course not: obstacles may block your line of vision, no matter from
which angle you look, making it impossible to determine the layout.
The central idea behind tomography is the, somewhat surprising, fact
that if you had performed absorption measurements from all angles,
instead of looking at shadows, you could indeed have found every
detail of the area.

Finding its first application in medicine, tomography techniques
have been used extensively in this field since the 1970's. Here,
tomography is used to reconstruct 3-D pictures from data recorded
using X-rays or NMR. 

With regard to quantum systems, tomography can be used to
reconstruct the density matrix of the system. For quantum harmonic
oscillator systems this reconstruction can be done exactly by
measuring the quadrature distributions for all quadrature angles;
for example measuring the density distribution in space for all
points of time during one period of oscillation. The techniques for
this include the inverse Radon transformation \cite{invradon},
\cite{invradon2} and the more reliable technique of pattern
functions \cite{patternfunc}. The analogy to the absorption
measurements mentioned earlier is apparent if one considers the
quadrature distributions as marginal distributions of the system's
Wigner function.

Buz\v ek and Drobn\'y have shown that a good approximation to the
density matrix for the one-particle harmonic oscillator can be
achieved using the MAXENT principle with the observations of just a
few quadrature distributions and the mean oscillator excitation
number \cite{Buzekjmo}. Skovsen et al. recently used MAXENT
tomography to reconstruct the one-body density matrix of free
particles from experimental data \cite{Juhl}.

In this paper we will extend the scope of MAXENT quantum tomography
by using it to approximately reconstruct the second moments of the
ladder operators, including the one-body density matrix, of a
many-particle system: a condensate treated in the Bogoliubov
approximation.

\subsection{\label{maxentteori}MAXENT principle for reconstruction of density operators}
The method we use to estimate the quantum state of our system at
hand is by finding the MAXENT density operator as suggested by
Jaynes \cite{jaynesto}. The goal of the method is to find a density
operator that exactly reproduces the knowledge one has about the
system while making as few assumptions as possible about degrees of
freedom one has no knowledge about. A proposal for this is the
maximum entropy density operator $\hat{\rho}_{ME}$ chosen for having
maximal von Neumann entropy $S = -Tr(\hat{\rho} \ln \hat{\rho} )$.
We make this more rigorous in the following way \footnote{The
statement of the MAXENT procedure in this paper is very similar to
the one given in f.x. \cite{Buzekatom}}:

Let a set of observables $\{ \hat{G}_\nu \}, (\nu \in {1,
2,\ldots,n})$ be associated with a quantum system prepared in an
unknown state $\hat{\rho}$. We imagine an ensemble of these quantum
systems wherein there are no correlations between the subsystems of
the ensemble, and we assume to have measured the ensemble
expectation values $\{ \bar{G}_\nu \}$ of the observables $\{
\hat{G}_\nu \}$.

Unless the set of observables $\{ \hat{G}_\nu \}$ constitutes the
\textit{quorum} (a complete set of observables) there will in
general be many density operators $\hat{\rho}_{\{\hat{G}\}}$ that
satisfy the normalization condition $Tr(\hat{\rho}_{\{\hat{G}\}}) =
1$ and predict the measured values:
\begin{eqnarray} \label{eqs:cond}
Tr(\hat{\rho}_{\{\hat{G}\}} \hat{G}_\nu) & = & \bar{G}_\nu ,\quad
\nu \in \{1,2,\ldots,n\}.
\end{eqnarray}

The MAXENT principle says that the most unbiased guess for the
density operator approximating $\hat{\rho}$ and fulfilling these
conditions is the one with maximal von Neumann entropy:
\begin{eqnarray}
S(\hat\rho_{ME}) &=& \max\big[S(\hat{\rho}_{\{\hat{G}\}}); \forall \hat{\rho}_{\{\hat{G}\}} \big], \\
S(\hat\rho) &=& -Tr(\hat{\rho} \ln \hat{\rho} ).
\end{eqnarray}
As shown by Jaynes \cite{jaynesto} this implies a density operator
of the form:
\begin{equation} \label{eq:rhome}
\hat\rho_{ME} = \frac{1}{Z_{\{\hat G\}}}\exp{\left(-\sum_{\nu=1}^n
\lambda_\nu\hat G_\nu \right)}
\end{equation}
with the generalized partition function:
\begin{equation} \label{eq:Tilstsum}
Z_{\{\hat G\}} = Tr\left[\exp{\left(-\sum_{\nu=1}^n \lambda_\nu\hat
G_\nu \right)}\right]
\end{equation}
and with the $\{\lambda_\nu\}$ being a set of Lagrange multipliers
chosen so that $\rho_{ME}$ in Eq.~(\ref{eq:rhome}) fulfills the
conditions Eqs.~(\ref{eqs:cond}). Using these $\lambda_\nu$'s we
have the explicit form of the MAXENT density operator
Eq.~(\ref{eq:rhome}).

A technical difficulty arises because, in all but the most trivial
circumstances, it is not feasible to analytically invert the
conditions Eqs.~(\ref{eqs:cond}) given Eq.~(\ref{eq:rhome}) to
obtain the set $\{ \lambda_\nu \}$. For implementation it is more
convenient, and we shall indeed use this procedure, to minimize the
norm of the differences between measured values and values predicted
by $\hat\rho_{ME}$ with respect to $\{\lambda_\nu\}$:
\begin{equation} \label{eq:fejlvektor}
\left( \begin{array}{c} Tr(\hat{\rho}_{ME}(\{\lambda_\nu\}) \hat{G}_1) - \bar{G}_1 \\
\vdots \\
Tr(\hat{\rho}_{ME}(\{\lambda_\nu\}) \hat{G}_n) - \bar{G}_n
\end{array} \right)
\end{equation}
Here, we have given the same weight to all components of the vector
of errors. One may wish to modify this if, for instance, some
observations are made with a much better precision than others.

\section{Description of the condensate}\label{sek:thecondensate}
\subsection{\label{bogteori} Bogoliubov Hamiltonian}
Let us consider a 3-dimensional cold gas of atoms having mass $m$,
occupying a volume V, and dominated by s-wave collisions where
$U(r,r') = U_{0}\delta(r-r')$ is the inter-particle potential. Such
a gas is well represented by the Hamiltonian:
\begin{eqnarray} \label{eq:fuldhamilton}
H_{full} &=& \sum_{k} \varepsilon_{k} \hat a^{\dagger}_{k} \hat
a_{k} + \frac{g}{2}\sum_{k, {k'}, q} \hat a^\dagger_{k+ q} \hat
a^\dagger_{{k'}-q} \hat a_{{k'}} \hat a_{k}
\end{eqnarray}
where $\varepsilon_k = \frac{\hbar^2k^2}{2m}$ is the free-particle
energy and $g = \frac{U_{0}}{V}$ is the coupling constant. The
operators $\hat a_k$ and $\hat a_k^\dagger$ are the usual boson
ladder operators annihilating and creating a particle with wave
vector $k$, and having the commutation relations:
\begin{eqnarray}
\lbrack \hat a_k , \hat a^\dagger_{k'} \rbrack & = & \delta_{k,k'} \label{eq:comm1}\\
\lbrack \hat a_k , \hat a_{k'} \rbrack & = & 0. \label{eq:comm2}
\end{eqnarray}
A Hamiltonian similar to Eq.~(\ref{eq:fuldhamilton}) applies in one
and two spatial dimensions with suitable redefinitions of $g$,
depending on the confinement of the remaining coordinates
\cite{1og2dimg}, \cite{1og2dimganden}. For simplicity, only one
spatial dimension will be used in this paper. The wave vectors $k$
are then merely scalars, and will in the following be referred to as
wave numbers.

Due to the complicated dynamics of the Hamiltonian
Eq.~(\ref{eq:fuldhamilton}), it is often necessary to use an
approximate Hamiltonian to make analytical calculations possible. We
will employ the so-called Bogoliubov approximation, which is valid
when the vast majority of the particles are in the zero wave number
mode \cite{Bogol}:
\begin{eqnarray} \label{eq:bogcond}
\langle \hat a^\dagger_0 \hat a_0 \rangle & \gg & \sum_{k \neq 0}
\langle \hat a^\dagger_k \hat a_k \rangle.
\end{eqnarray}
Having this condition fulfilled, it is a good approximation to
approximate the Hamiltonian Eq.~(\ref{eq:fuldhamilton}) by
\footnote{Going directly from Eq.~(\ref{eq:fuldhamilton}), we have
subtracted a constant, having no physical significance}:
\begin{eqnarray} \label{eq:Boghamilton}
\hat H &=& \frac{1}{2}\sum_{k\neq 0}\bigg(\big(\varepsilon_k+g
N_{tot} \big)
(\hat a^\dagger_k \hat a_k +\hat a_k \hat a^\dagger_k)+ \nonumber \\
&& \qquad \qquad + \, g N_{tot}\big(\hat a_k \hat a_{-k} + \hat
a^\dagger_{-k} \hat a^\dagger_{k}\big) \bigg)
\end{eqnarray}
and $N_{tot}$ is the total number of particles in the gas. This
operator is bilinear in the ladder operators, and can be
diagonalized by a change of basis. We will do this by using
elementary linear algebra.

\subsection{Discretization of position and momentum}
We will now specify some useful technical details of our treatment.
We consider an odd number $(N > 1)$ of evenly spaced points in 1-D
coordinate space, symmetrically distributed around (and including)
the origin, on an interval of length $L = N\Delta x$.
\begin{eqnarray} \label{eq:nvaerdier}
x_n &=& n \Delta x, \qquad n = -M, -M+1,\ldots,M
\end{eqnarray}
where $M = \frac{N-1}{2}$. We assume that the spatial atomic density
is measured on this grid, i.e., our observations will be of the form
$\langle \hat \psi^\dagger(x_n,t) \hat \psi(x_n,t) \rangle$. Due to
the form of the Hamiltonian Eq.~(\ref{eq:Boghamilton}), it is
convenient to introduce the discrete wave numbers corresponding to
the spatial discretization above.
\begin{eqnarray} \label{eq:boelgetal}
k_q &=& q \Delta k, \qquad \Delta k = \frac{2 \pi}{L},
\end{eqnarray}
where $q$ can assume the same values as $n$ above.

The usual discrete Fourier transform from coordinate space ($ \hat
\psi(x_n) \text{ and } \hat \psi^\dagger(x_n) $) to wave number
space ($\hat a(k_q) \text{ and } \hat a^\dagger(k_q) $) reads:
\begin{eqnarray} \label{eq:almfourierf}
\hat a(k_q) &=& \frac{1}{\sqrt N}\sum^{M}_{n = -M}\hat
\psi(x_n)\exp{(ik_qx_n)} \\
\hat a^\dagger(k_q) &=& \frac{1}{\sqrt N}\sum^{M}_{n = -M}\hat
\psi^\dagger(x_n)\exp{(-ik_qx_n)}. \label{eq:almfourierb}
\end{eqnarray}

All the ladder operators in coordinate space can be arranged in a
$2N\times 1$ column vector:
\begin{eqnarray} \label{eq:psiopvektor}
\bm \psi &=& \left(
\begin{array}{c}
  \hat \psi(x_{-M}) \\
  \vdots \\
  \hat \psi(x_{M}) \\
  \hat \psi ^\dagger(x_{-M}) \\
  \vdots \\
  \hat \psi ^\dagger(x_{M})
\end{array}
\right),
\end{eqnarray}
and correspondingly:
\begin{equation}
\bm \psi^\dagger = \left(\hat
\psi^\dagger(x_{-M}),\ldots,\hat\psi^\dagger(x_{M}),\hat\psi(x_{-M}),\ldots,\hat\psi(x_{M})\right).
\end{equation}
In this paper, we shall denote vectors of this type by bold letters.
Similarly, we can construct the column vector $\bm a$ from the
operators in wave number space $\hat a(k_q)$ and $\hat
a^\dagger(k_q)$:
\begin{eqnarray} \label{eq:aopvektor}
\bm a &=& \left(
\begin{array}{c}
  \hat a(k_{-M}) \\
  \vdots \\
  \hat a(k_{M}) \\
  \hat a^\dagger(k_{-M}) \\
  \vdots \\
  \hat a^\dagger(k_{M})
\end{array}
\right),
\end{eqnarray}
Using these vectors, we can write the Fourier transformation
Eqs.~(\ref{eq:almfourierf})--(\ref{eq:almfourierb}) as a matrix
multiplication by a $2N\times 2N$ matrix:
\begin{equation} \label{eq:fouriermat}
\bm a = \mathscr A \cdot \bm \psi, \quad \text{whereby} \quad \bm
a^\dagger = \bm \psi^\dagger \cdot \mathscr A^\dagger .
\end{equation}
Here, and in the following, we shall use script letters to denote
matrices of dimension $2N\times 2N$.

The matrix $\mathscr A$ has the following form:
\begin{equation}
\mathscr A =\left(%
\begin{array}{cc}
  F & 0 \\
  0 & F^* \\
\end{array}%
\right)
\end{equation}
and $F$ is the usual $N\times N$ discrete Fourier transformation
matrix, whose $q$'th row is:
\begin{equation}
F(q,:) = \frac{1}{\sqrt
N}\left[\exp{(ik_qx_{-M})},\ldots,\exp{(ik_qx_M)}\right]
\end{equation}
from which Eqs.~(\ref{eq:almfourierf})--(\ref{eq:almfourierb}) are
easily recovered.

It is straightforward to see that the operators $\hat a(k_q)$ and
$\hat\psi(x_n)$ obey similar commutation relations
Eqs.~(\ref{eq:comm1})-(\ref{eq:comm2}). In this paper we shall
exclusively deal with transformations to new sets of operators that
conserve this property.

\subsection{Diagonalization of the Bogoliubov Hamiltonian}
Using the notation developed in the preceding section, we can write
the Bogoliubov Hamiltonian Eq.~(\ref{eq:Boghamilton}) in a compact
form:
\begin{equation}
\hat H = \bm a^\dagger \mathscr H \bm a .
\end{equation}
As shown in e.g. \cite{Colpa}, this Hamiltonian can be diagonalized
to the matrix $\mathscr E$ by transforming to new boson operators:
\begin{eqnarray}
\bm b &=& \mathscr B \bm a \label{eq:bogmat}\\
\nonumber \\
\hat H &=& \bm a^\dagger \mathscr H \bm a \nonumber \\
 &=& \bm a^\dagger
\mathscr{B^\dagger (B^\dagger)}^{-1} \mathscr{H B}^{-1} \mathscr B \bm a \nonumber \\
 &=& \bm b^\dagger \mathscr{(B^\dagger)}^{-1} \mathscr{H B}^{-1} \bm
 b \nonumber \\
 &=& \bm b^\dagger \mathscr E \bm b, \\
\nonumber \\
\mathscr E &=& \left(\mathscr{B}^\dagger \right)^{-1} \mathscr{H
B}^{-1}. \label{eq:bogtransf}
\end{eqnarray}
The matrix $\mathscr H$ only has elements different from zero in the
two diagonals $\mathscr H(k_q,k_q)$ and $\mathscr H(k_q,k_{-q})$,
giving the matrix an "X-structure". In addition, all elements are
real. Therefore the matrix $\mathscr B^{-1}$ that diagonalizes
$\mathscr H$ also only has elements different from zero with these
indices, and can be chosen real. The upper-left to lower-right
diagonal elements are denoted $u(k_q)$, while the upper-right to
lower-left diagonal elements are denoted $v(k_q)$. The elements can
be found in e.g. \cite{Stringari}:
\begin{displaymath}
u(k_q) = \left\{ \begin{array}{ll}
\qquad 1 & \text{for }  q = 0 \\
\sqrt{\frac{q^2/4 + \gamma/2}{\epsilon(q)/E_0} +\frac{1}{2}} &
\text{for }  q \neq 0
\end{array} \right.
\end{displaymath}
\begin{displaymath}
v(k_q) = \left\{ \begin{array}{ll} 0 & \text{for } q = 0 \\
- \sqrt{\frac{q^2/4 + \gamma/2}{\epsilon(q)/E_0}-\frac{1}{2}} &
\text{for }  q \neq 0 .
\end{array} \right.
\end{displaymath}

The diagonalized form $\mathscr E$ has elements $\epsilon(k_q)$:
\begin{eqnarray}
 \epsilon(k_q) &=& E_0\sqrt{\gamma q^2 + q^4/4}
\end{eqnarray}
where $E_0 = \frac{(2\pi\hbar)^2}{mL^2}$, $\gamma =
\frac{gN_{tot}}{E_0}$, and $gN_{tot}$ has the same meaning as in
Eq.~(\ref{eq:Boghamilton}).

 By diagonalizing the Hamiltonian we have found a basis of
non-interacting modes with a simple time evolution:
\begin{equation}
\hat b_q(t) = \hat b_q(0)\exp{(-i\epsilon(q)t/\hbar)}. \nonumber
\end{equation}
By defining the para-identity matrix $\mathscr{\hat I}$, which will
play an important role in the next section:
\begin{equation} \label{eq:paraid}
\mathscr{\hat I} =
diag(\underbrace{1,\ldots,1,}_{N}\underbrace{-1,\ldots,-1}_{N})
\end{equation}
we can write the vector of operators $\bm b$ at time $t$ as:
\begin{equation} \label{eq:bbasis}
\bm b(t) = \exp{(-i\mathscr{\hat I E}t/\hbar)} \bm b(0) =
\mathscr{U}(t) \bm b(0).
\end{equation}

The changes of basis can be summarized like this:
\begin{eqnarray}\label{eq:psitilb}
\bm \psi(t) &=& \mathscr A^{-1} \bm a(t) \nonumber \\
 &=& \mathscr A^{-1} \mathscr B^{-1} \bm b(t) \nonumber \\
 &=& \mathscr A^{-1} \mathscr B^{-1} \mathscr U(t) \bm b(0) ,
\end{eqnarray}
whereby we can find the time evolution of the $\hat \psi(x_n,t)$ and
$\hat \psi^\dagger(x_n,t)$:
\begin{equation}\label{eq:psitilpsi}
\bm \psi(t) = \mathscr A^{-1}\mathscr{B}^{-1}\mathscr U(t)
\mathscr{BA}\bm \psi(0).
\end{equation}

\subsection{General para-unitary diagonalization}
All the vectors of operators Eq.~(\ref{eq:psiopvektor}),
Eq.~(\ref{eq:aopvektor}) and Eq.~(\ref{eq:bogmat}) are examples of a
vector of general boson ladder operators:
\begin{eqnarray} \label{eq:genopvektor}
\bm \alpha &=& \left(
\begin{array}{c}
  \hat \alpha_{-M} \\
  \vdots \\
  \hat \alpha_{M} \\
  \hat \alpha^\dagger_{-M} \\
  \vdots \\
  \hat \alpha^\dagger_{M}
\end{array}
\right),
\end{eqnarray}
fulfilling the commutation relations:
\begin{eqnarray}
\lbrack \hat \alpha_q , \hat \alpha^\dagger_{q'} \rbrack & = & \delta_{q,q'} \label{eq:gencomm1}\\
\lbrack \hat \alpha_q , \hat \alpha_{q'} \rbrack & = & 0
\label{eq:gencomm2}
\end{eqnarray}
Both the Fourier transformation Eq.~(\ref{eq:fouriermat}) and the
Bogoliubov transformation Eq.~(\ref{eq:bogmat}) are examples of
changes of basis, known as para-unitary transformations, which
conserve the Boson commutation relations between the operators. All
the transformations to new sets of operators in this paper conserve
this property. In our present treatment, there are two main reasons
for making these transformations. Firstly, the measurements we have
performed on the system are in the coordinate space and wave number
space at different times, but the operators corresponding to these
measurements have complicated time evolutions. Working instead in
the $\bm b$-basis Eq.~(\ref{eq:bbasis}) the time evolution is
simple, and the exponent in Eq.~(\ref{eq:rhome}) can be easily
calculated. Secondly, to perform the traces in
Eq.~(\ref{eq:fejlvektor}), it is convenient to shift to a basis
where the density operator is diagonal, and this can be done by a
para-unitary transformation. We recall that there are $2N$ ladder
operators, regardless of basis (f.x. the set $\{\hat \psi(x_n)\}
\text{ and } \{\hat \psi^\dagger(x_n)\}$), which will make the
transformation matrices $2N \times 2N$.

The commutation relations demand that in a transformation from one
set of boson operators to another:
\begin{equation}
\bm \beta = \mathscr T \bm \alpha
\end{equation}
the $2N\times 2N$ matrix $\mathscr T$ must be para-unitary, i.e.
satisfy the condition \footnote{To follow \cite{Colpa} we use the,
in matrix algebra unusual, prefix "para-" instead of f.x.
"pseudo-".}:
\begin{eqnarray}
\mathscr{\hat I} &=& \mathscr{T \hat I T^\dagger}
\end{eqnarray}
with the diagonal matrix $\mathscr{\hat I}$ defined in
Eq.~(\ref{eq:paraid}) \footnote{Had we been treating fermion-
instead of boson-operators, the matrix $\mathscr T$ would have been
unitary.}. From the definition it is seen that the product of two
para-unitary matrices is again para-unitary.

The matrices we will diagonalize in this paper will all be
hermitian, positive definite and have the structure \footnote{An
bilinear, hermitian operator can always be brought to this form by
adding a scalar constant which is determined by the diagonal of the
matrix.}:
\begin{eqnarray} \label{eq:genmatstruk}
\mathscr X =\left(
\begin{array}{cc}
  P & Q \\
  Q^* & P^* \\
\end{array}
\right) .
\end{eqnarray}
Let $\hat X$ be a hermitian operator in the $\bm \alpha$ basis with
coefficient matrix $\mathscr X$. Such a hermitian matrix $\mathscr
X$ can be diagonalized by the para-unitary matrix $\mathscr T$ to a
new set of operators $\{\beta_n\}$ and $\{\beta^\dagger_n\}$:
\begin{eqnarray}
\hat X &=& \bm \alpha^\dagger \mathscr X \bm \alpha \\
 &=& \bm \alpha^\dagger
\mathscr{T^\dagger (T^\dagger)}^{-1} \mathscr{X T}^{-1} \mathscr T \bm \alpha \nonumber \\
 &=& \bm \beta^\dagger \mathscr{(T^\dagger)}^{-1} \mathscr{X T}^{-1} \bm
 \beta \nonumber \\
 &=& \bm \beta^\dagger \mathscr L \bm \beta .
\end{eqnarray}
Because $\mathscr X$ is positive definite the para-eigenvalues
$\mathscr L_{i,i}$ will also be positive by Sylvester's law of
inertia. The matrix $\mathscr T$ can be chosen to have the
structure:
\begin{equation} \label{eq:gentransstruk}
\mathscr T = \left(
\begin{array}{cc}
  U & V^* \\
  V & U^* \\
\end{array}
\right)
\end{equation}
giving $\mathscr L$ the form:
\begin{equation}
\mathscr L = diag(\mathscr L_{1,1}, \mathscr L_{2,2},\ldots,\mathscr
L_{N,N},\mathscr L_{1,1},\ldots,\mathscr L_{N,N}) .
\end{equation}

This general procedure will become useful in the following section,
where we will need to evaluate expectation values of the observed
densities within a quantum state of the form Eq.~(\ref{eq:rhome}),
facilitated by a para-unitary diagonalization of the exponent
$\sum_{\nu}\lambda_\nu \hat G_\nu$.

\section{\label{sek:minteori}Using MAXENT with the Bogoliubov approximation}
Now we are ready to use the MAXENT formalism on the condensate in
the Bogoliubov approximation. As the set of observables $\{\hat
G_\nu\}$ we use measurements of particle numbers at coordinate space
points at different times $t$, and of the occupation of wave number
modes at times $t'$. The times $t$ can, but need not, be equal to
the times $t'$
\begin{eqnarray}
\{\hat G_\nu\} &=& \{\hat \psi^\dagger(x_n,t) \hat \psi(x_{n},t) + \hat \psi(x_n,t) \hat \psi^\dagger(x_{n},t)\} \bigcup \nonumber \\
&& \{\hat a^\dagger(k_q,t')\hat a(k_q,t')+\hat a(k_q,t')\hat
a^\dagger(k_q,t')\} .
\end{eqnarray}
The reason for using this symmetrical form of the observables is
that it leads to the favorable structure Eq.~(\ref{eq:genmatstruk})
of the matrix in the exponent of Eq.~(\ref{eq:rhome}).

The task is now to put these observables into the MAXENT density
operator Eq.~(\ref{eq:rhome}) and perform traces like
Eqs.~(\ref{eqs:cond}). To do this we will use the same method as in
section \ref{teori} to diagonalize the matrix in the exponent into a
set of new, non-interacting quasi-particles.

Let $\mathscr W_n$ be the $2N\times 2N$ matrix with the following
elements:
\begin{equation}
\mathscr W_n(r,s) = \left\{ \begin{array}{ll}
 1 & \quad \text{for } r=s=n+M+1 \\
 1 & \quad \text{for } r=s=n+N+M+1\\
 0 & \quad \text{otherwise}
\end{array} \right.
\end{equation}
so that (using Eq.~(\ref{eq:psitilb}) for the first part):
\begin{equation}
\hat \psi^\dagger(x_n,t) \hat \psi(x_{n},t) + \hat \psi(x_n,t) \hat
\psi^\dagger(x_{n},t) = \bm \psi^\dagger(t) \mathscr W_n \bm \psi(t)
\nonumber
\end{equation}
\begin{equation}
= \bm b^\dagger(0) \mathscr U^\dagger(t) \left(\mathscr B^{-1}\right)^\dagger \left(\mathscr A^{-1}\right)^\dagger \mathscr W_n \mathscr A^{-1} \mathscr B^{-1} \mathscr U(t) \bm b(0) \label{eq:psiop} \\
\end{equation}
\\
\begin{equation}
\hat a^\dagger(k_q,t')\hat a(k_q,t')+\hat a(k_q,t')\hat a^\dagger(k_q,t') = \bm a^\dagger(t') \mathscr W_q \bm a(t') \nonumber \\
\end{equation}
\begin{equation}
= \bm b^\dagger(0) \mathscr U^\dagger(t') \left(\mathscr
B^{-1}\right)^\dagger \mathscr W_q \mathscr{B}^{-1} \mathscr U(t')
\bm b(0). \label{eq:aop}
\end{equation}

The operator $\sum_{\nu}\lambda_\nu \hat G_\nu$, with $\hat G_\nu$
being operators of the form Eq.~(\ref{eq:psiop}) and
Eq.~(\ref{eq:aop}) is hence expressed in terms of the Bogoliubov
eigenmode operators, and we can write the MAXENT density operator
Eq.~(\ref{eq:rhome}) in a compact form:
\begin{eqnarray} \label{eq:rhomemin}
\hat \rho_{ME} &=& \frac{1}{Z}\exp{(-\bm b^\dagger(0) \mathscr P \bm
b(0))},
\end{eqnarray}
where
\begin{eqnarray}
\mathscr P &=& \sum_t \sum_{n=-M}^{M} \lambda(n,t) \cdot \nonumber \\
 && \qquad \quad \mathscr U^\dagger(t) \left(\mathscr B^{-1}\right)^\dagger \left(\mathscr A^{-1}\right)^\dagger \mathscr W_n \mathscr A^{-1} \mathscr B^{-1} \mathscr U(t) \nonumber \\
&&+ \sum_{t'} \sum_{q=-M}^{M} \lambda(q,t') \cdot \nonumber \\
&& \qquad \quad \mathscr U^\dagger(t') \left(\mathscr
B^{-1}\right)^\dagger \mathscr W_q \mathscr{B}^{-1} \mathscr U(t') .
\label{eq:Pdef}
\end{eqnarray}
Since the structures of $\mathscr U(t)$, $\mathscr A$ and $\mathscr
B$ are all like that of Eq.~(\ref{eq:gentransstruk}), and the
structure of $\mathscr W_q$ is like that of
Eq.~(\ref{eq:genmatstruk}), so is the structure of $\mathscr P$. We
can therefore diagonalize it to a diagonal matrix $\mathscr L$ by a
para-unitary change of basis to new Boson operators $\bm c =
\mathscr C \bm b(0)$:
\begin{eqnarray}
\bm b^\dagger(0) \mathscr P \bm b(0) &=& \bm c^\dagger \mathscr L
\bm c \\
&=& \sum^{N}_{j=1}\mathscr L_{j,j}\big(\hat c^\dagger_j\hat c_j+\hat
c_j\hat c^\dagger_j\big).
\end{eqnarray}
Remembering that the $\mathscr L_{j,j}$ are all positive, the
partition function Eq.~(\ref{eq:Tilstsum}) becomes:
\begin{eqnarray}
Z &=& Tr\left(\exp\left[-\sum^{N}_{j=1}\mathscr L_{j,j}\left(\hat
c^\dagger_j\hat c_j+\hat c_j\hat c^\dagger_j\right)\right]\right) \nonumber \\
&=& \prod_{j=1}^{N}\left(\frac{1}{\exp{(\mathscr
L_{jj})}-\exp{(-\mathscr L_{jj})}}\right), \label{eq:Z}
\end{eqnarray}
and we can readily determine the corresponding set of expectation
values:
\begin{eqnarray}
\langle \hat c^\dagger_m \hat c_n \rangle &=& \frac{1}{Z}
Tr\left(\hat c^\dagger_m \hat c_n\exp\left[\sum^{N}_{j=1}\mathscr
L_{j,j}\left(\hat c^\dagger_j\hat c_j+\hat c_j\hat
c^\dagger_j\right)\right]\right) \nonumber \\
&=& -\frac{1}{2}\left(1+\frac{\partial \ln Z}{\partial \mathscr
L_{n,n}}\right)\delta_{m,n} \nonumber \\
&=& \frac{1}{\exp(2\mathscr L_{n,n})-1)}\delta_{m,n} \qquad \text{and}\\
\nonumber \\
\langle \hat c_m \hat c_n \rangle &=& \langle \hat c^\dagger_m \hat
c^\dagger_n \rangle = 0.
\end{eqnarray}
Knowing these expectation values, it is easy to find the expectation
values of all the second moments in any basis, for example:
\begin{eqnarray}
\langle \bm \psi(t) \bm \psi^\dagger(t) \rangle &=& \nonumber
\end{eqnarray}
\begin{equation}
\left(
\begin{array}{cccc}
  \langle \hat \psi(x_{-M}) \hat \psi^\dagger(x_{-M}) \rangle & \cdots & \langle \hat \psi(x_{-M}) \hat \psi(x_{-M}) \rangle & \cdots \\
  \vdots & \ddots & \vdots & \ddots \\
  \langle \hat \psi^\dagger(x_{-M}) \hat \psi^\dagger(x_{-M}) \rangle & \cdots & \langle \hat \psi^\dagger(x_{-M}) \hat \psi(x_{-M}) \rangle & \cdots \\
  \vdots & \ddots & \vdots & \ddots \\
\end{array} \right) \label{eq:forvpsikpsi} \nonumber \\
\end{equation}
\begin{eqnarray}
&=& \mathscr A^{-1} \langle \bm a(t) \bm a^\dagger(t) \rangle \left(\mathscr A^{-1}\right)^\dagger \nonumber \\
&=& \left[\mathscr{CU}^{-1}(t)\mathscr{BA}\right]^{-1} \langle \bm c \bm c^\dagger \rangle \cdot \nonumber \\
& & \left\{\left[\mathscr {CU}^{-1}(t)\mathscr{BA}\right]^{-1}
\right\}^\dagger. \nonumber
\end{eqnarray}
In passing, it is noted that the lower right $N\times N$ submatrix
of the matrix $\langle \bm \psi(t) \bm \psi^\dagger(t) \rangle$ is
the one-body density matrix.

As a final point we give a comment on the requirement of the
positive definiteness of $\mathscr P$. It is clearly seen here that
we must require the diagonal elements of $\mathscr L$ to be strictly
positive for the geometrical series summed to find Eq.~(\ref{eq:Z})
to be convergent. Therefore, by Sylvester's law of inertia,
$\mathscr P$ must also be positive definite. In addition, for this
condition to be fulfilled, the number of observables in $\{\hat
G_\nu\}$ must be at least as large as N. The reason for this is that
in Eq.~(\ref{eq:Pdef}) the matrices $\mathscr W_n$ have
dimensionality two in the $2N$ space, the dimensionality of course
being conserved under the para-unitary changes of basis. So to span
the $2N$ space, at least N observables of dimensionality two must be
included. In the present paper, however, we will have many more
observations than N, and linear dependencies between them are highly
unlikely to reduce the dimensionality below $2N$.

\section{Tomography on three quantum states\label{sek:eksemplerne}}

\subsection{Initial excited states}\label{sek:tilstande}
The time evolution is given by Eq.~(\ref{eq:psitilpsi}). Since we
are not attempting to reconstruct the full many-particle quantum
state, but only the second moments of the ladder operators, it is
hence sufficient to specify these at, for example, $t = 0$:
\begin{eqnarray}
\langle \bm \psi(t) \bm \psi^\dagger(t) \rangle &=& \nonumber \\
&&\mathscr A^{-1} \mathscr B^{-1}\mathscr{U}(t)\mathscr{BA} \langle \bm \psi(0) \bm \psi^\dagger(0)\rangle \cdot \nonumber \\
&& \left[\mathscr A^{-1}\mathscr B^{-1}\mathscr{U}(t)
\mathscr{BA}\right]^\dagger . \label{eq:startbet}
\end{eqnarray}

To examine the reliability of the MAXENT technique, regarding
reconstruction of the correct second moments of the ladder
operators, we study three very different types of condensate quantum
states with nearly the same density distribution at $t = 0$: namely
a gaussian perturbation on top of a flat condensate.

\begin{enumerate}
\item \textbf{Coherent state} is a pure state of all the particles, representable as a product state with all the particles in the same state.
We choose this to be a constant with a gaussian perturbation in
coordinate space.
\begin{eqnarray}
|State \, 1\rangle &=& \big(\hat \psi^\dagger_{g} \big)^{N_{part}} |0\rangle, \quad\text{where} \\
 \hat \psi^\dagger_{g} &=&
 \sum_{x_n}\phi_{gauss}(x_n)\hat \psi^\dagger(x_n)
\end{eqnarray}
and $\phi_{gauss}$ is a constant plus a gaussian with standard
deviation $\sigma$, centered at the origin: $\phi_{gauss} \propto
\left\{1+\eta
 \exp\left[-\frac{1}{2}\left(\frac{x_n}{\sigma}\right)^2\right]\right\}$. $N_{part}$ is the
total number of particles. We can envision this state formed in a
uniform system with a negative potential dip, having $State \, 1$ as
its ground state. The spatial evolution of the state after abruptly
turning off the potential at $t = 0$ is what is usually handled with
the Gross-Pitaevskii equation:
\begin{equation}\label{eq:gpdiff}
-i\hbar \frac{\partial \psi}{\partial t} = \left(-
\frac{\hbar^2}{2m}\frac{\partial^2}{\partial x^2} + g
N_{tot}|\psi|^2 \right) \psi .
\end{equation}
\item \textbf{Thermal perturbation}
This state is a flat condensate with a thermal gaussian perturbation
superposed hereupon. The gaussian perturbation is treated as
originating from a thermal boson gas in a harmonic trap and we have
assumed no initial correlations between the perturbation and the
original flat condensate. Therefore the second moments for this
state in coordinate space are simply the sum for the thermal
gaussian and the flat condensate.
\item \textbf{Squeezed condensate} We get this state by applying an
operator, similar to the squeezing operator known from squeezing of
light, to a condensate with all $N_{part}$ particles in the
condensate mode:
\begin{equation}
|State \, 3\rangle = \exp\left[½z^*(\hat \psi_{g})^2-z(\hat
\psi_{g}^\dagger)^2\right]\left(\hat b_0^\dagger
\right)^{N_{part}}|0\rangle .
\end{equation}
The most important difference between the two preceding states and
this one is the magnitude of the anomalous second moments at $t =
0$. In $State \, 1$ and $2$ the anomalous second moments are all
zero, where they in this state are comparable in magnitude with the
normal second moments.
\end{enumerate}

\subsection{\label{sek:resultater}Results of tomography}
We present here the results of using the MAXENT technique on the
three states discussed in section \ref{sek:tilstande}. The main
points of interest are not just whether a reliable reconstruction is
attained, but also the amount of data needed for this. In all the
results shown in this section we have used density distributions in
coordinate space at four different times. In $State \, 2$ we have
furthermore used the momentum distribution at $t = 0$, and in $State
\, 3$ we have additionally included an observation of $\langle \hat
a_0 \hat a_0 + \hat a_0^\dagger \hat a_0^\dagger \rangle$.

We have used $10^5$ particles in the flat part of the condensate at
$t = 0$ and set $g N_{tot} = 0.1$. The population in the zero wave
number modes is 99.5\% in $State \, 1$ and $3$ and 96\% in $State \,
2$. For the calculations we have used a grid of $N = 25$ discrete
points.

\begin{figure}[htbp]
\includegraphics[width=0.4\textwidth]{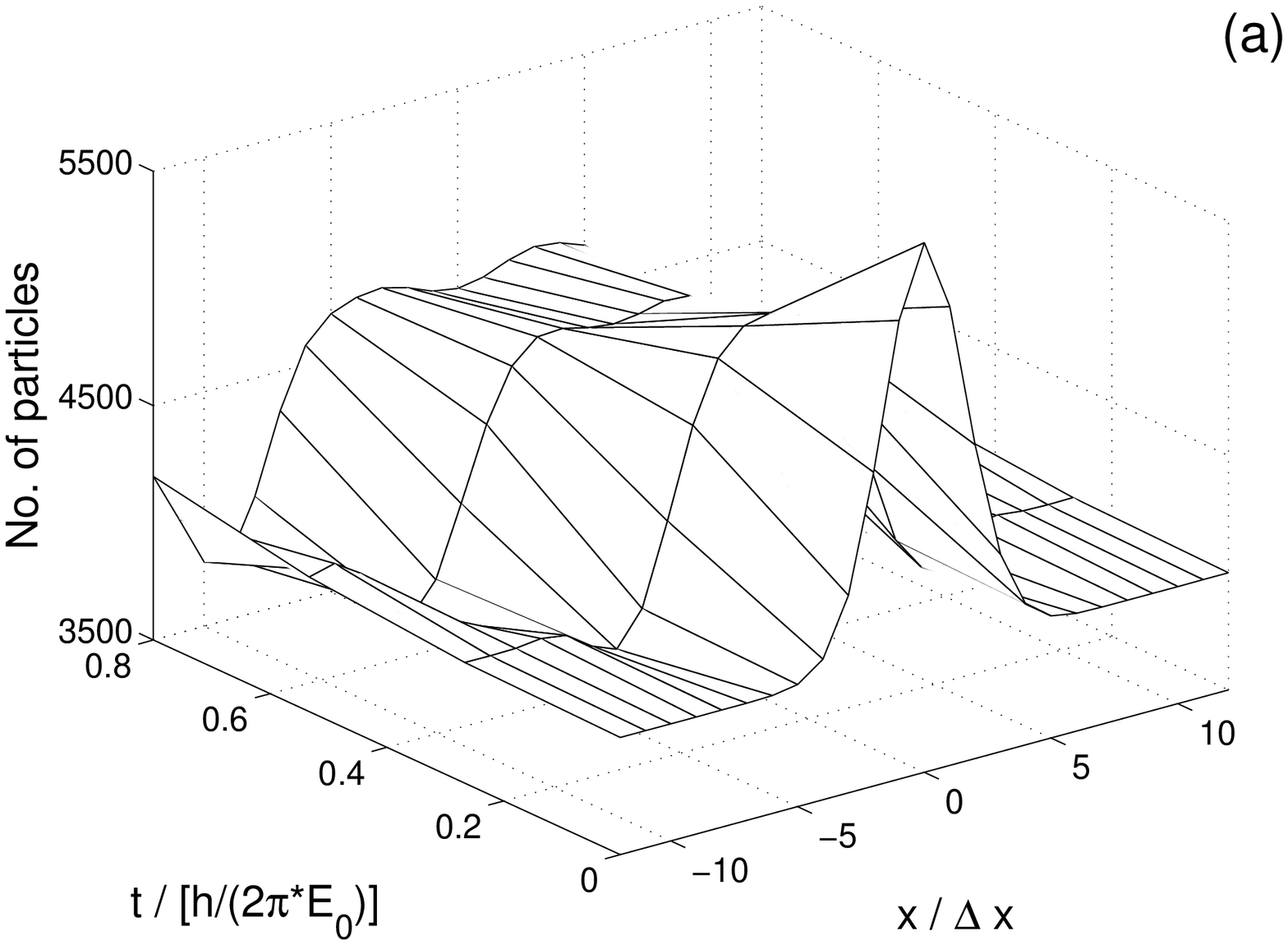}
\includegraphics[width=0.4\textwidth]{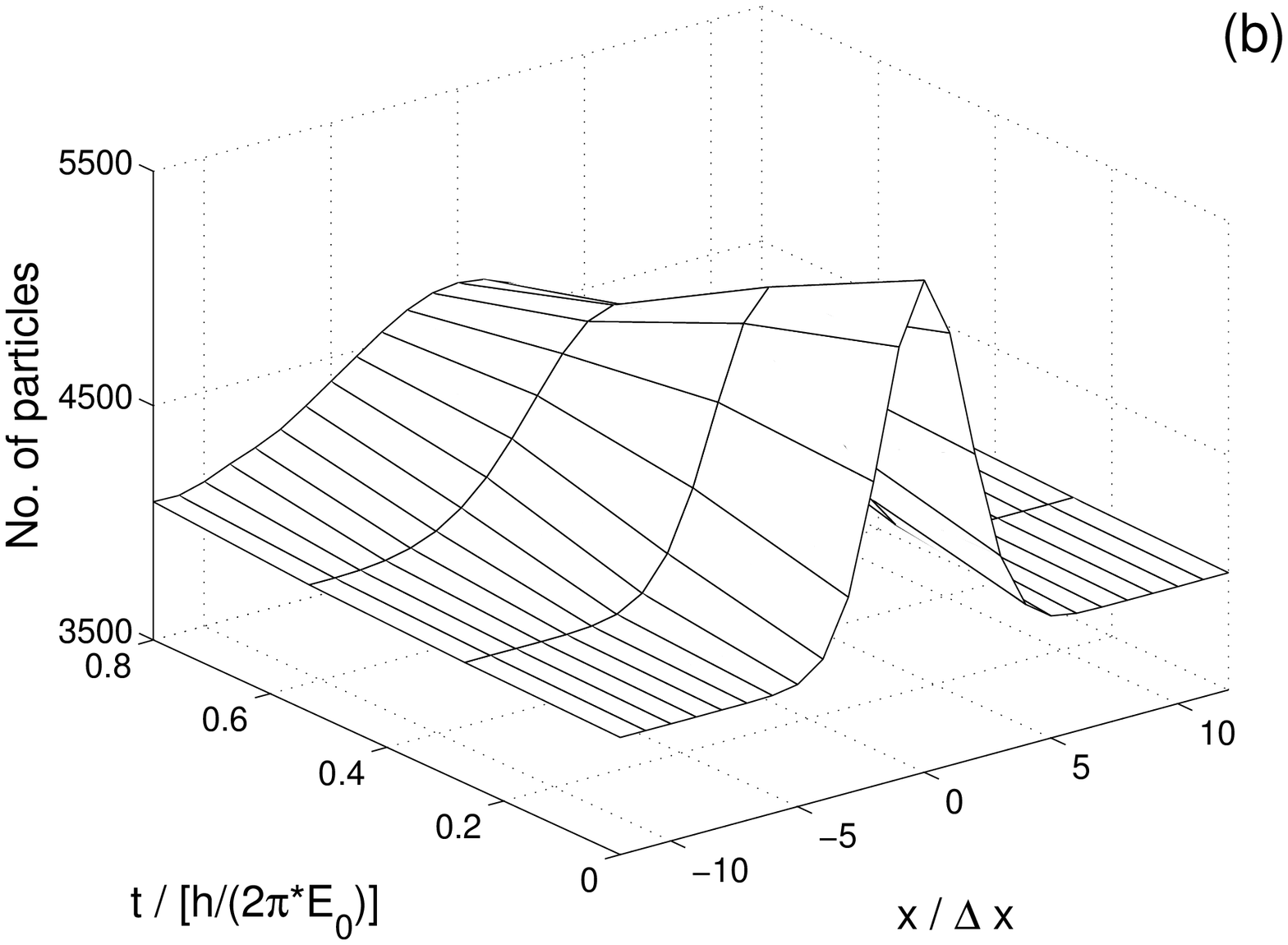}
\includegraphics[width=0.4\textwidth]{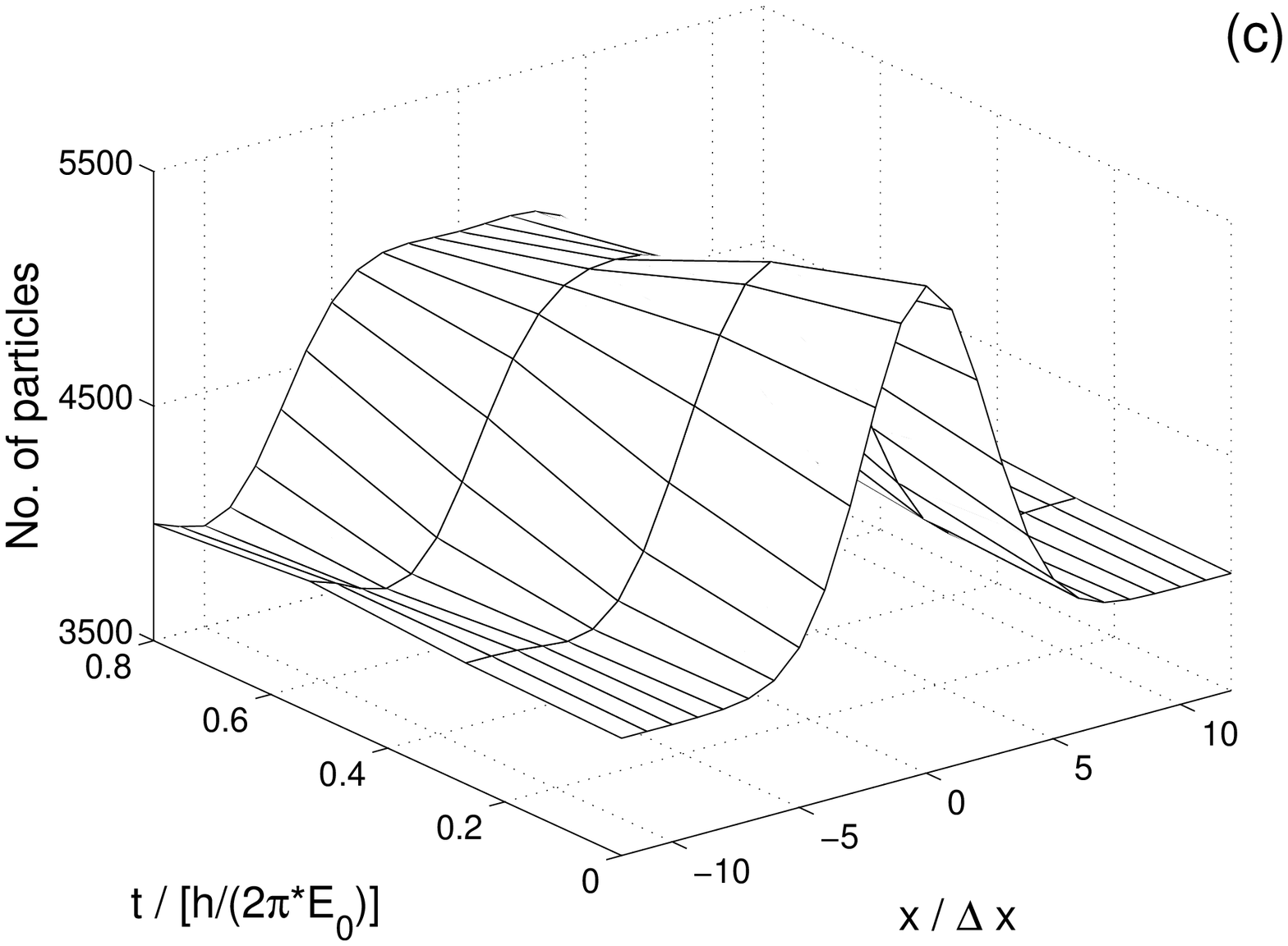}
\caption{\label{fig:stedd}Density distributions in coordinate space
at four different points of time for: (a) $State \, 1$, (b) $State
\, 2$ and (c) $State \, 3$. The data shown are those used in the
calculations together with momentum distributions at $t = 0$ for
$State \, 2$ and $3$, and one anomalous moment for $State \, 3$.}
\end{figure}

In Fig.~\ref{fig:stedd} we show the density distributions in
coordinate space of the three different states declared in section
\ref{sek:tilstande}. We have intentionally chosen the initial
distributions to be very similar. The perturbations give rise to
density variations of a magnitude readily detected \cite{Ketterle1}.

We now proceed to examine the true second moments of the ladder
operators for the three states and those reconstructed using the
MAXENT procedure. We will study these using surface plots of the
absolute value of the second moments and of the absolute value of
the difference between the true values of these moments and those
predicted by MAXENT for the state at $t = 0$. In this way, errors in
both magnitude and phase of the matrix elements will be visible.

For $State \, 1$, where we will see that the reconstruction is
precise and only the normal second moments are non-zero at $t = 0$,
we shall also display the phase-space Wigner function corresponding
to these normal second moments. Apart from normalization, the normal
second moments can be interpreted as the density matrix for any
single atom in the system, and the phase space Wigner function
W$(x_n,k_q)$ is a convenient representation of this. The cited works
\cite{Buzekjmo}, \cite{Buzekatom}, \cite{Juhl} all show the Wigner
function of the particle states, and much interest has been devoted
to the fact that measurements of exclusively positive marginal
distributions may be used to identify Wigner functions with domains
of negative values \cite{Wineland}, \cite{Lvovsky}. The connection
between the Wigner function and the normal second moments is given
in appendix \ref{app:wignersjov}.

\subsubsection{State 1} 
\begin{figure}[htbp]
\includegraphics[width = 0.5\textwidth]{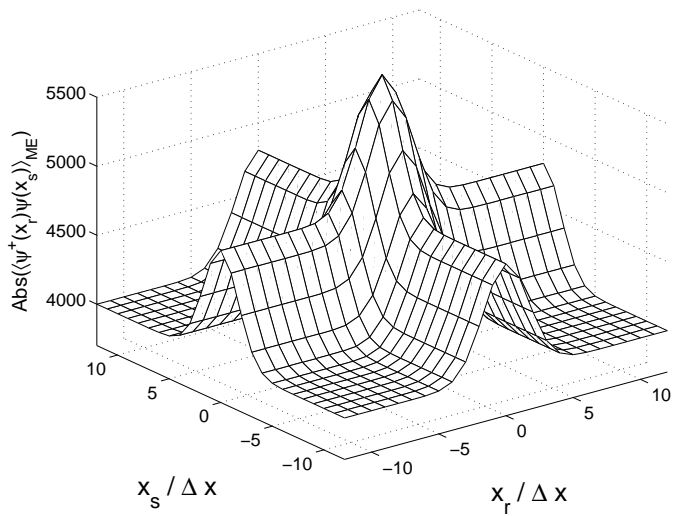}
\caption{Absolute value of the reconstructed normal second moments
$\langle \hat \psi^\dagger(x_r) \hat \psi(x_s) \rangle$ of $State \,
1$. The anomalous second moments are all zero (not shown). Both
normal and anomalous moments are reconstructed to within machine
precision, including complex phase, giving accumulated errors of
magnitude $10^{-4}$ on each element. The data used for the
reconstruction are four spatial density distributions. The true
second moments are not shown, as they are practically identical to
the reconstructed ones.\label{fig:2mom1}}
\end{figure}

Treating first $State \, 1$,
Fig.~\ref{fig:2mom1} presents the absolute value of the normal
second moments as found from section \ref{sek:tilstande}. The
reconstructed second moments are identical to the input values
within roundoff errors, which amounts to an accumulated error of
$10^{-4}$ on each matrix element, including phase. The same is true
for the anomalous second moments, all having the true value zero at
$t = 0$.

\begin{figure}[htbp]
\includegraphics[width = 0.5\textwidth]{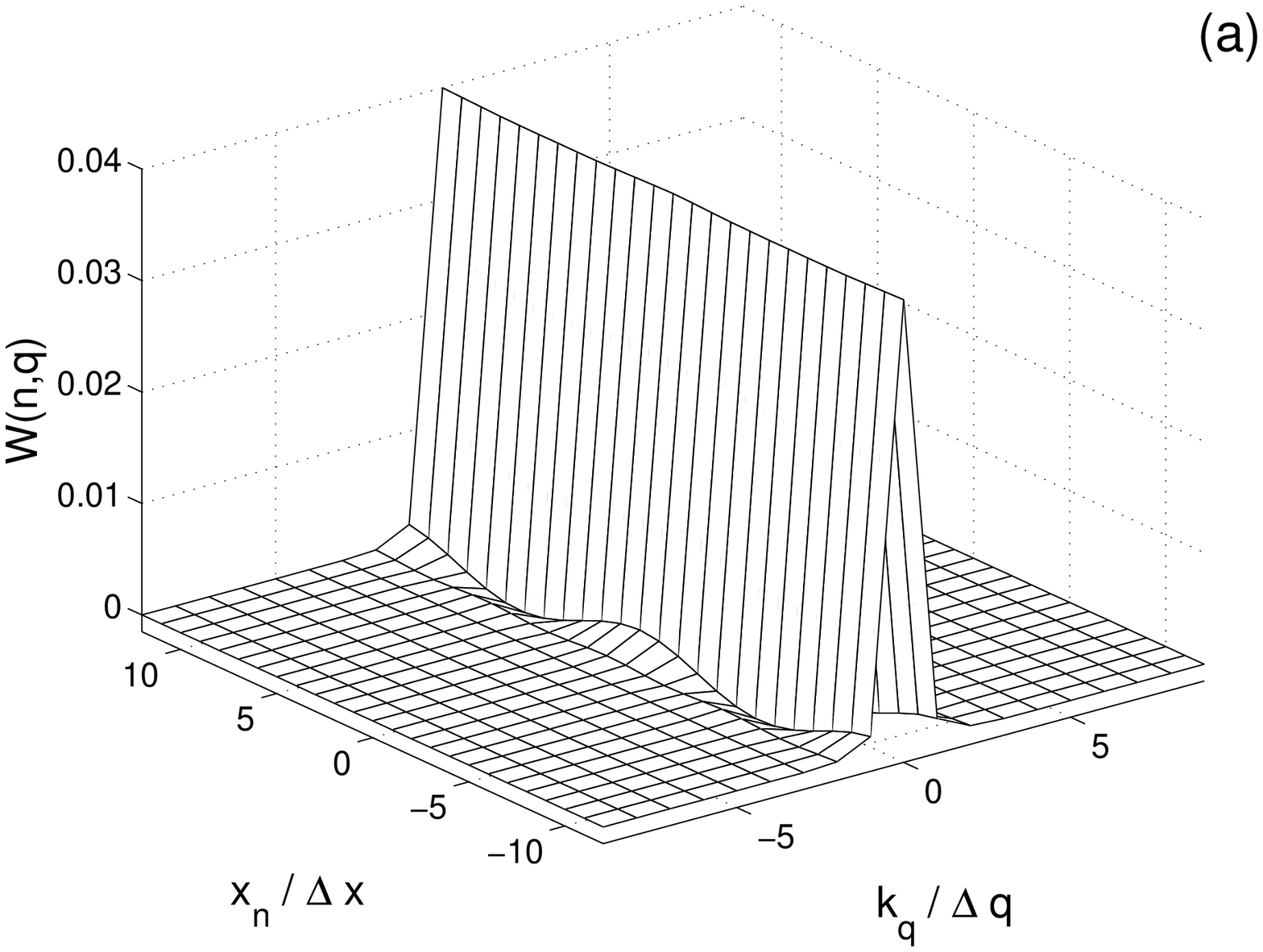}
\\
\includegraphics[width = 0.5\textwidth]{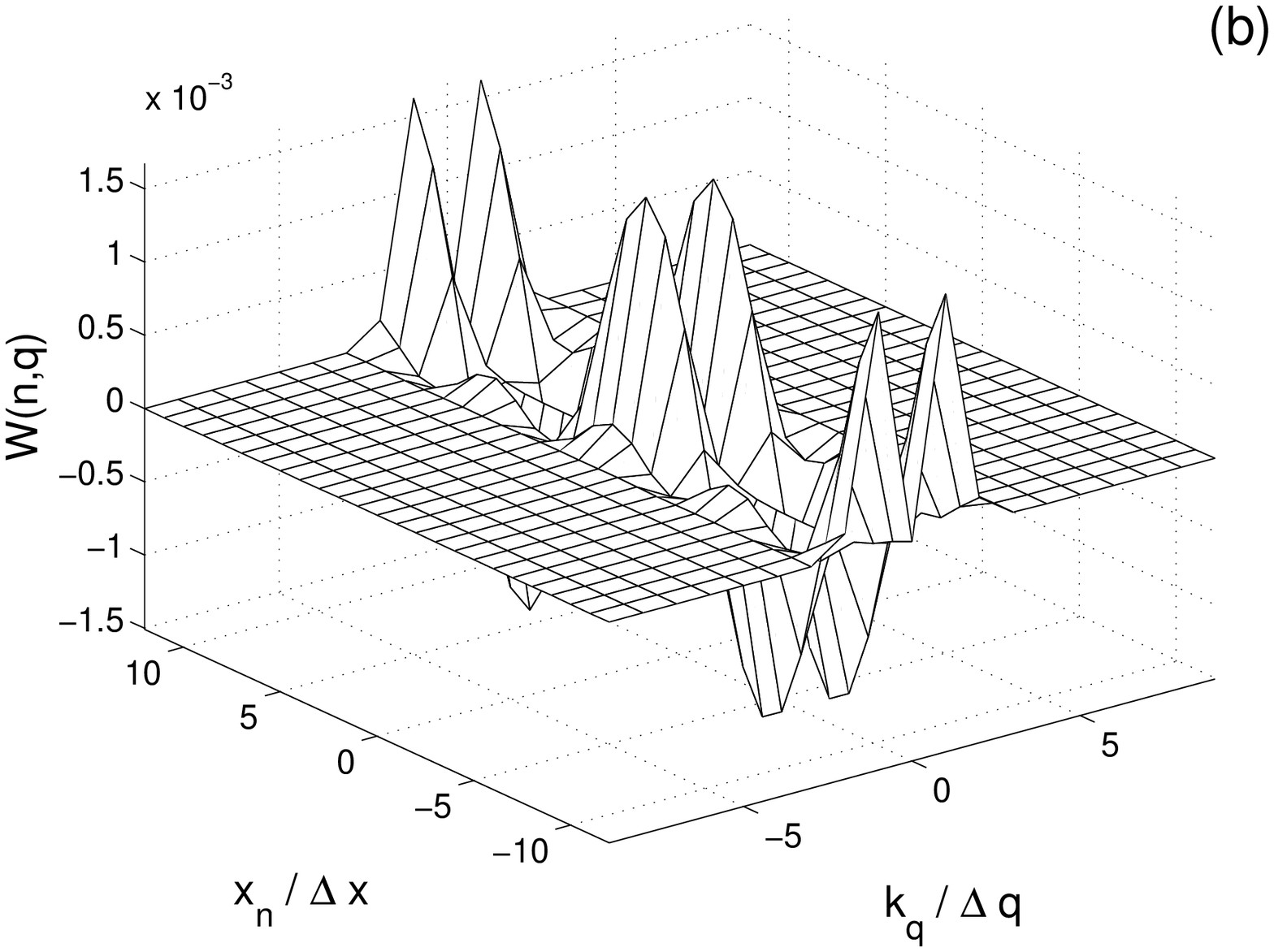}
\caption{The upper graph (a) shows the Wigner function for the
normal second moments of $State \, 1$, i.e. the one-body density
matrix apart from normalization. In the lower graph (b) the elements
W(n,0) are set equal to zero to more clearly reveal the finer
structures and the negativities of the Wigner function.
\label{fig:wigner}}
\end{figure}

As a consequence, the Wigner function for $State \, 1$ is also
perfectly reconstructed, and is presented in Fig.~\ref{fig:wigner}.

\subsubsection{State 2} 
\begin{figure}[htbp]
\includegraphics[width = 0.5\textwidth]{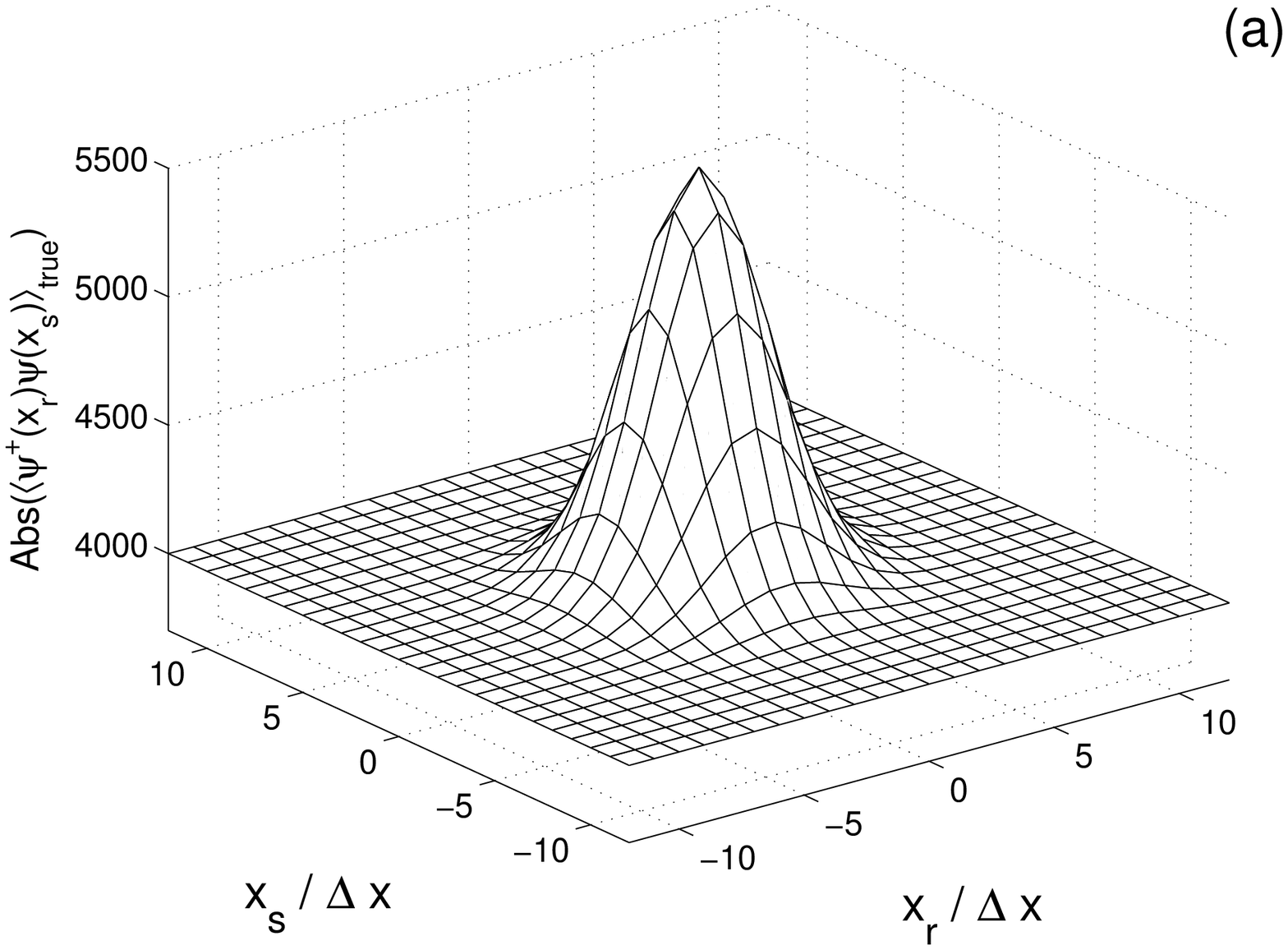}
\includegraphics[width = 0.5\textwidth]{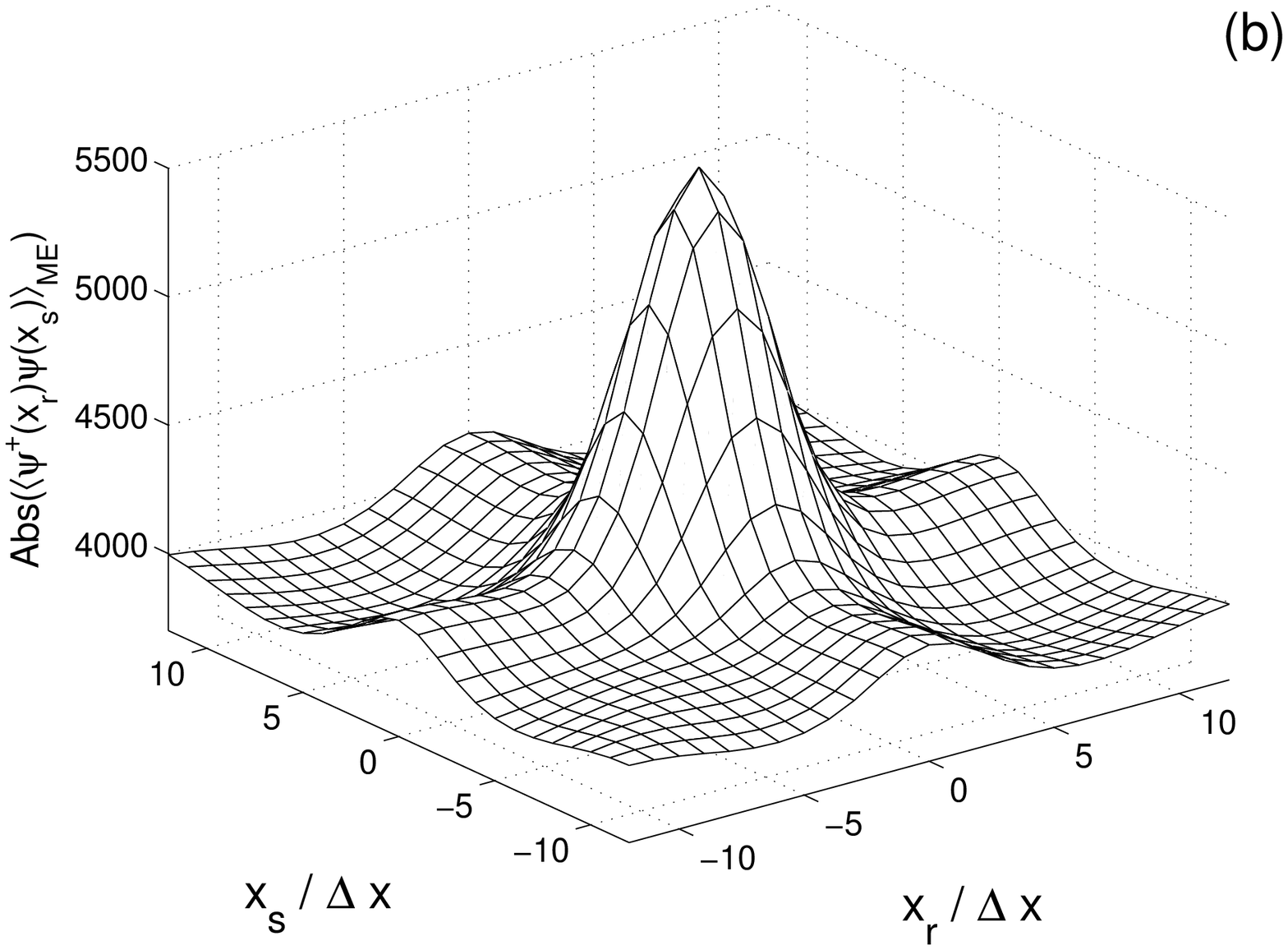}
\includegraphics[width = 0.5\textwidth]{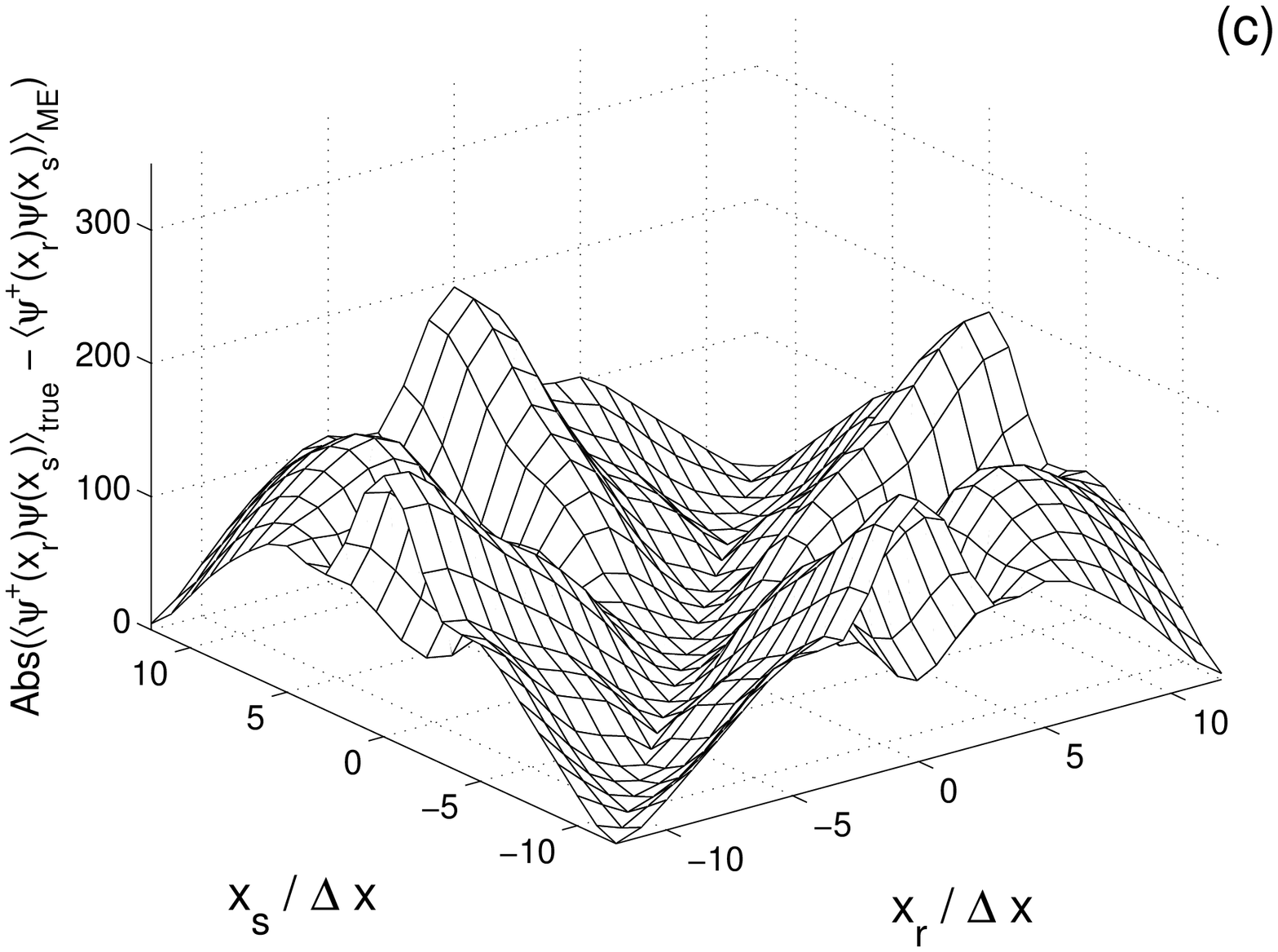}
\caption{\label{fig:st2nmom}The first two graphs show the absolute
value of (a) the true, and (b) the reconstructed normal second
moments $\langle \hat \psi^\dagger(x_r) \hat \psi(x_s) \rangle$ for
$State \, 2$. The errors above the condensate are up to about 10\%
of the peak value. The bottom graph (c) shows the absolute value of
the difference between true and reconstructed normal second moments.
The data used in the reconstruction are four spatial density
distributions and one momentum distribution.}
\end{figure}

For $State \, 2$ the true and reconstructed
normal second moments are shown in Fig.~\ref{fig:st2nmom}. Also
shown in this figure is the absolute value of the difference between
the true and reconstructed normal second moments. The errors are
many orders of magnitude larger than for $State \, 1$.

\begin{figure}[htbp]
\includegraphics[width = 0.5\textwidth]{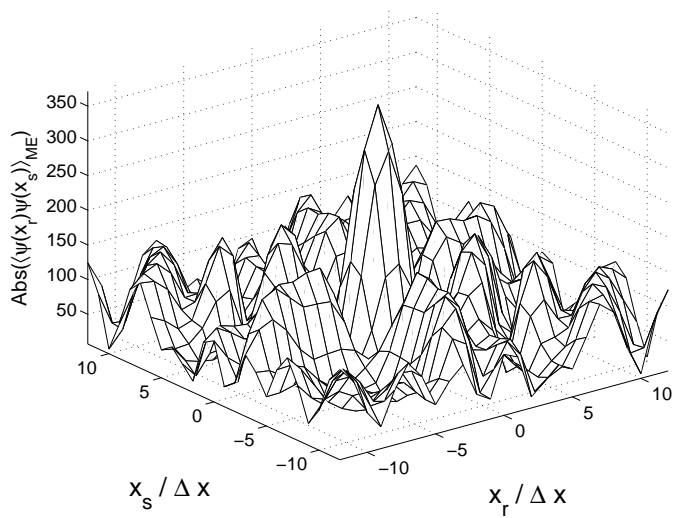}
\caption{Absolute value of the reconstructed anomalous second
moments $\langle \hat \psi(x_r) \hat \psi(x_s) \rangle$ for $State
\, 2$ at $t = 0$. The true values are all zero at $t = 0$. The
magnitude of the errors is similar to the errors of the normal
second moments for this state.\label{fig:st2nogadiff}}
\end{figure}

In Fig.~\ref{fig:st2nogadiff} we display the absolute value of the
reconstructed anomalous second moments, that all have the true value
zero at $t = 0$.

\subsubsection{State 3} 
\begin{figure}[htbp]
\includegraphics[width=0.5\textwidth]{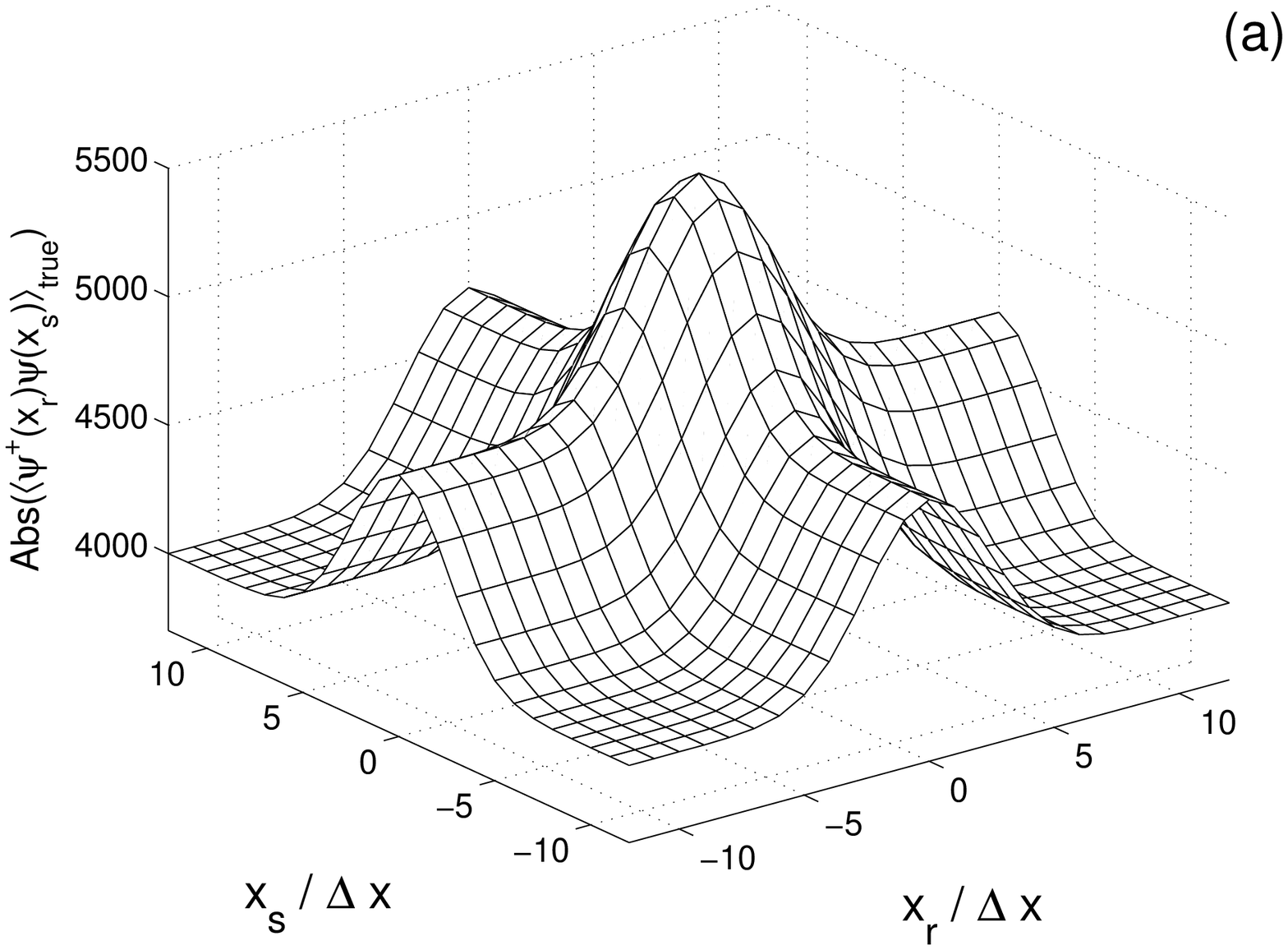}
\includegraphics[width=0.5\textwidth]{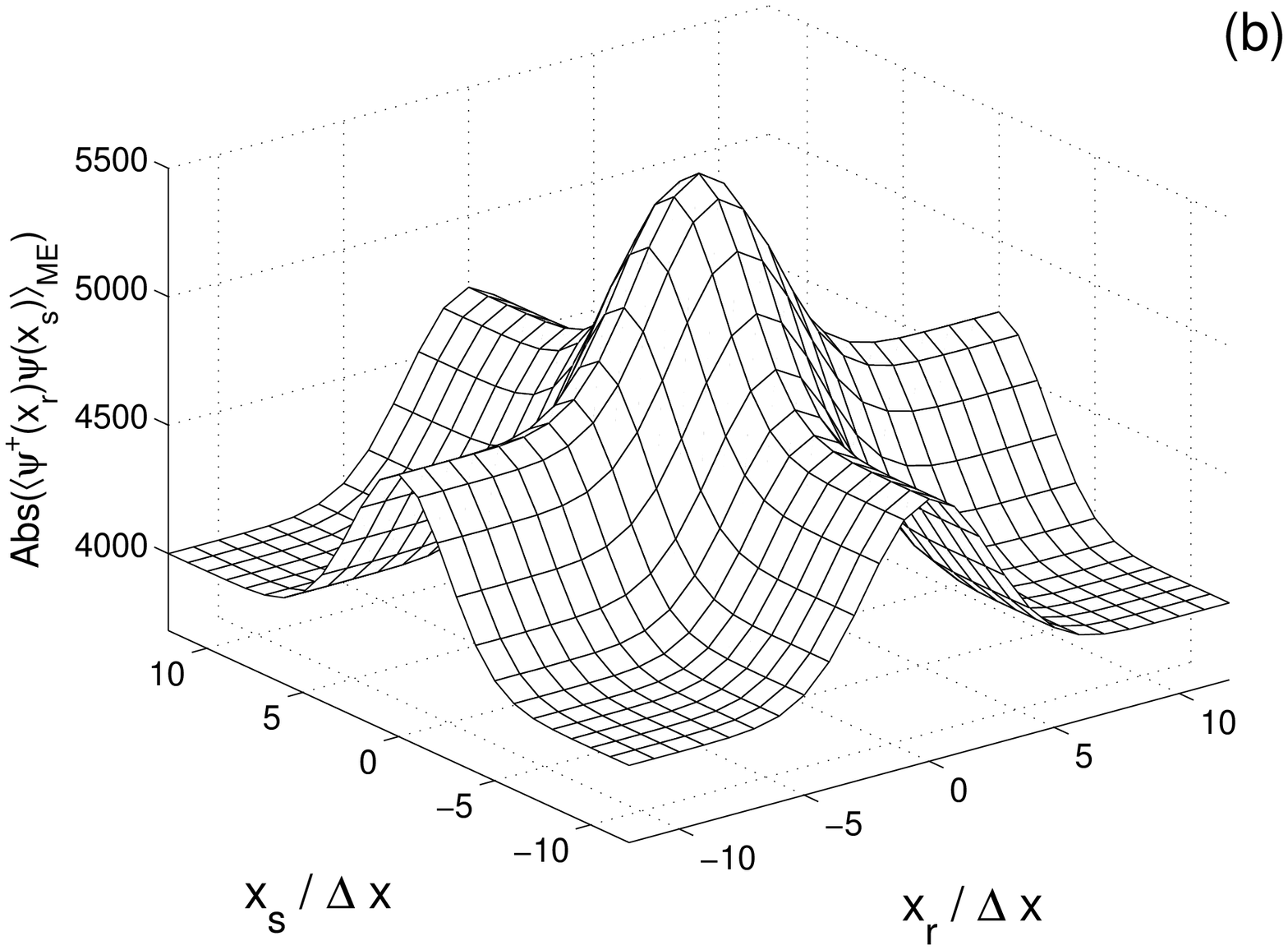}
\includegraphics[width=0.5\textwidth]{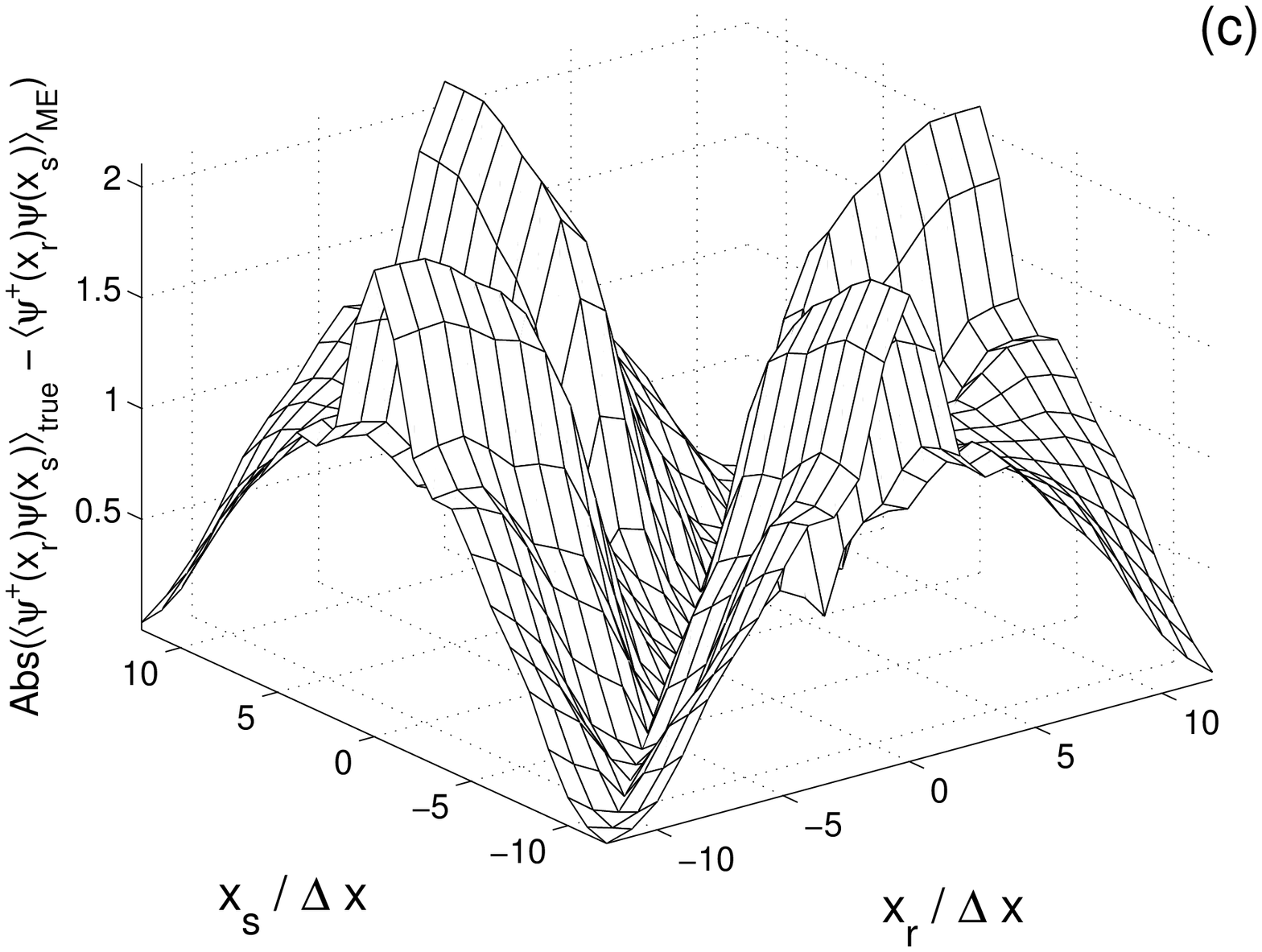}
\caption{\label{fig:st3nmom}The first two graphs show the absolute
value of (a) the true, and (b) the reconstructed normal second
moments $\langle \hat \psi^\dagger(x_r) \hat \psi(x_s) \rangle$ of
$State \, 3$. The lowest graph (c) is the absolute value of the
difference between the true and reconstructed second moments. The
data used in the reconstruction are four spatial density
distributions, one momentum distribution and $\langle \hat a_0 \hat
a_0 + \hat a_0^\dagger \hat a_0^\dagger \rangle$.}
\end{figure}

The absolute value of the true and reconstructed second moments are shown in
Fig.~\ref{fig:st3nmom} together with the absolute value of the
difference between these.

\begin{figure}[htbp]
\includegraphics[width = 0.5\textwidth]{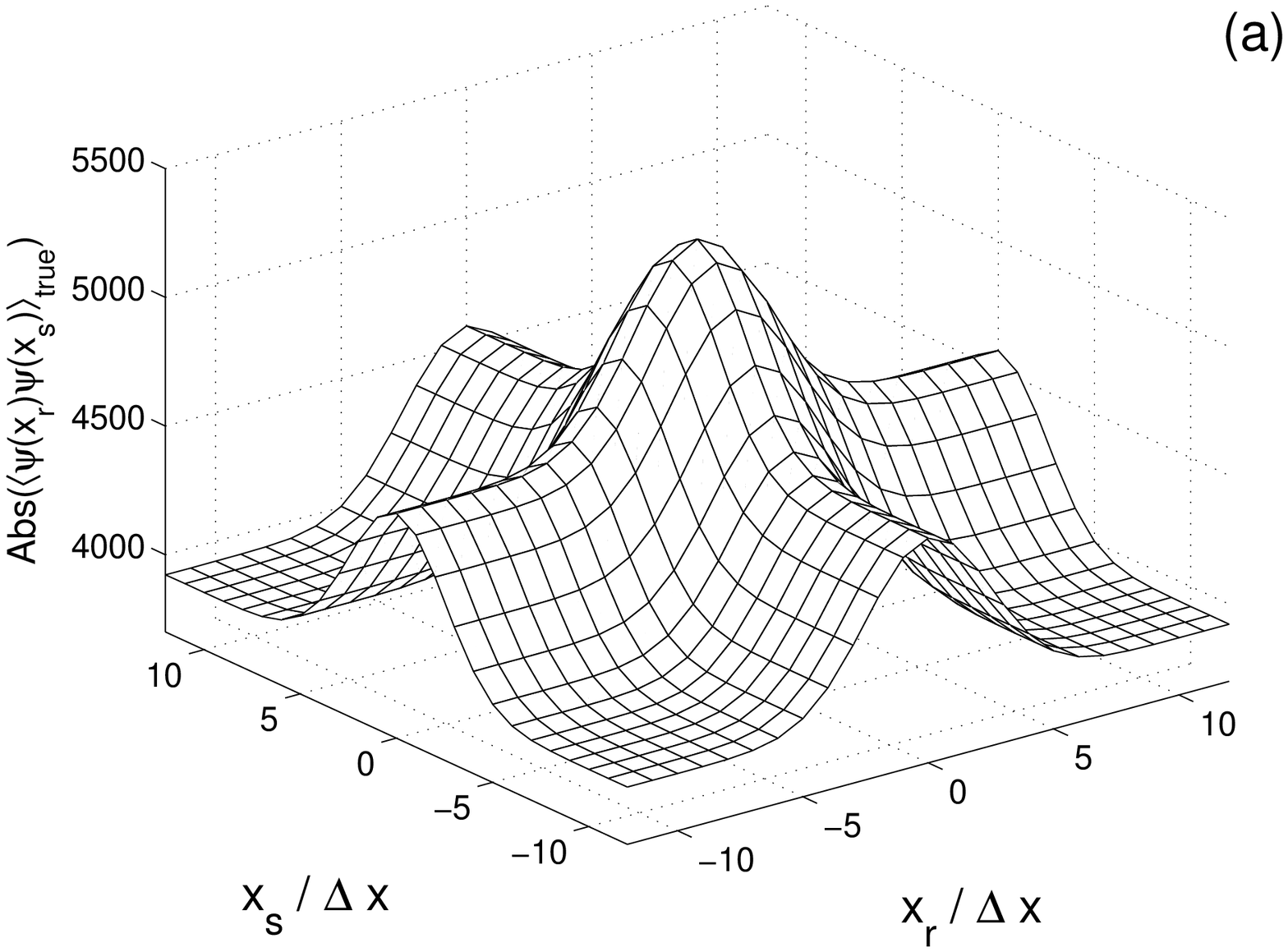}
\includegraphics[width = 0.5\textwidth]{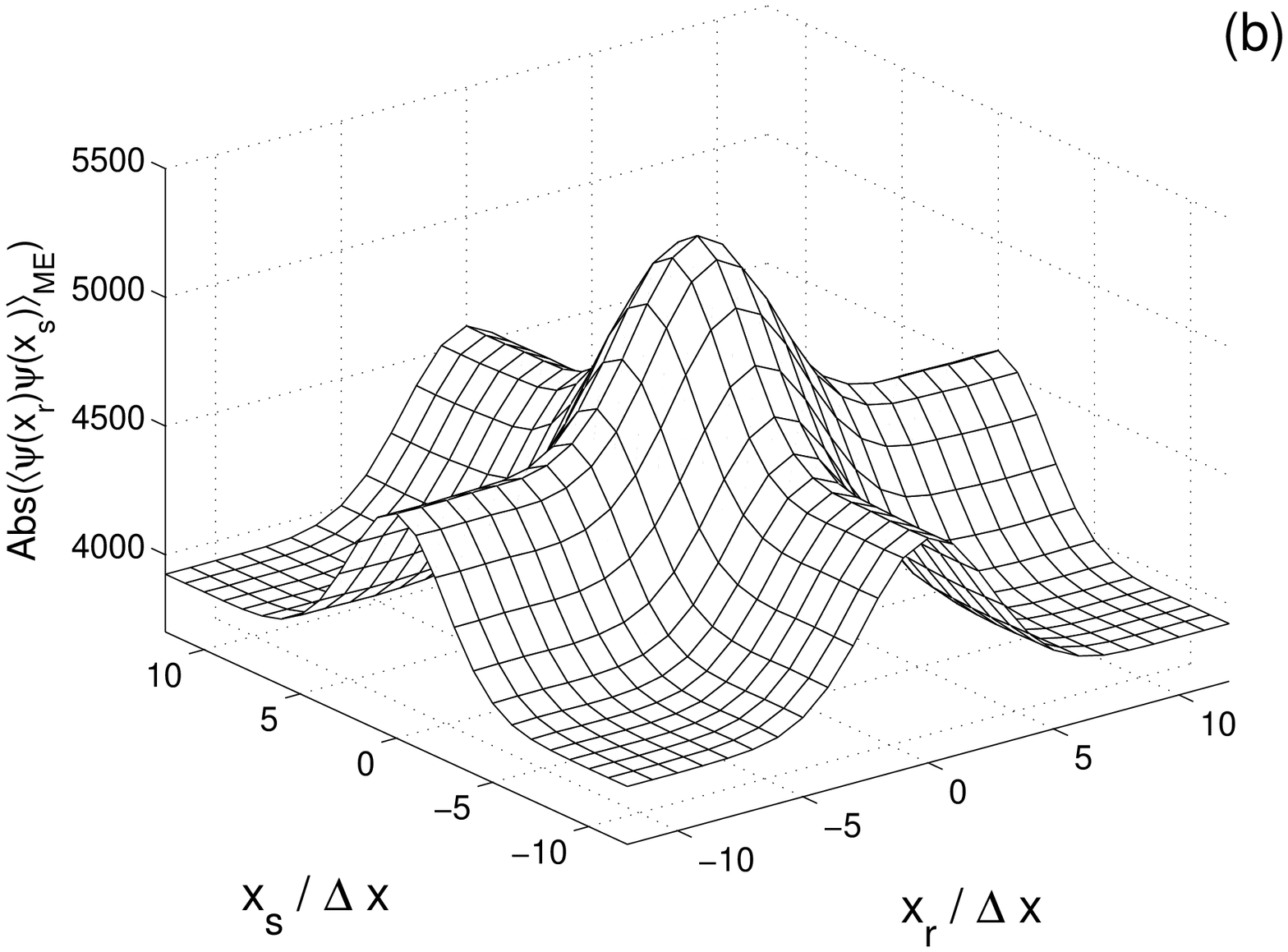}
\includegraphics[width = 0.5\textwidth]{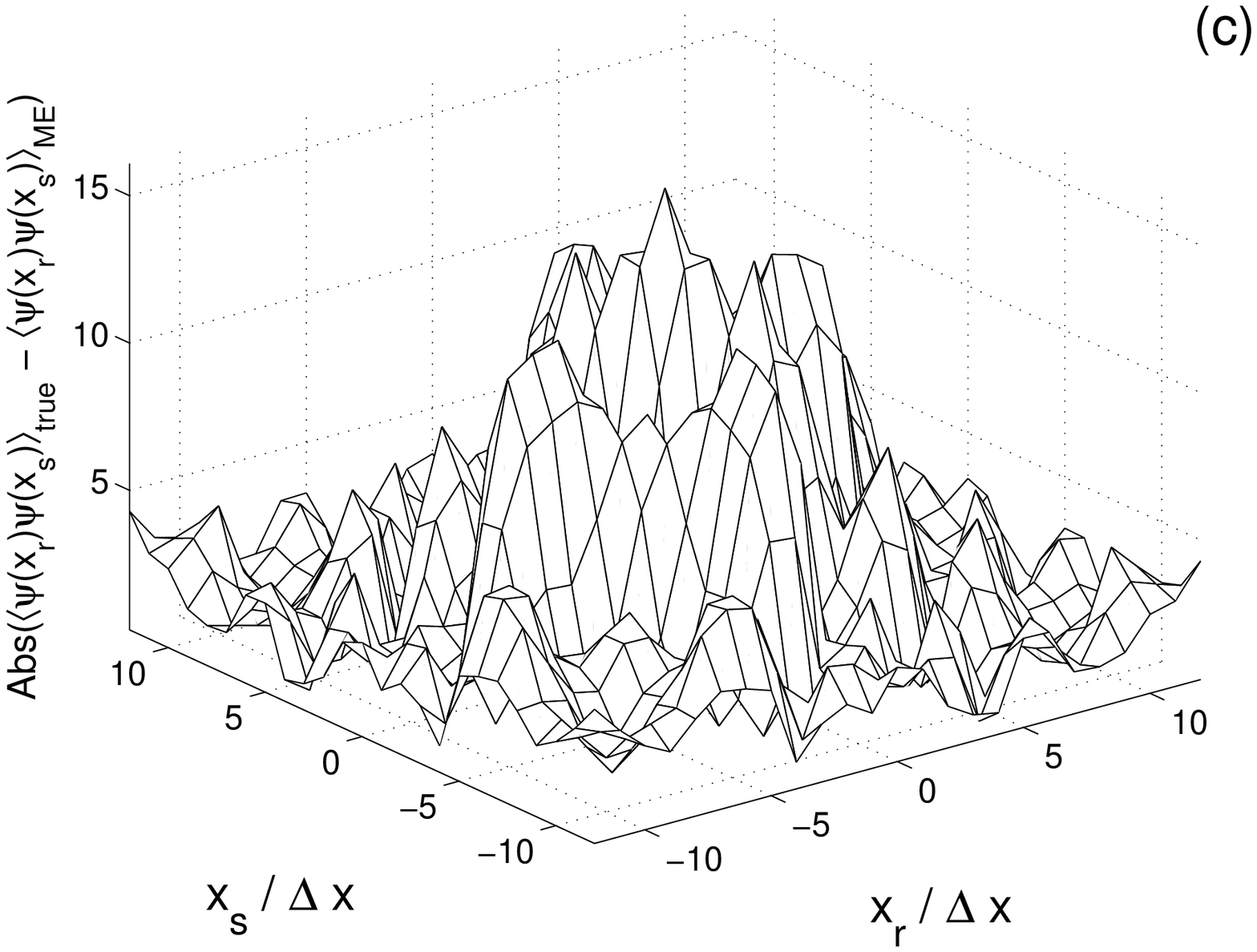}
\caption{The first two graphs show the absolute value of (a) the
true, and (b) the reconstructed anomalous second moments $\langle
\hat \psi(x_r) \hat \psi(x_s) \rangle$ of $State \, 3$. The lowest
graph (c) is the absolute value of the difference between the true
and reconstructed second moments. The data used in the
reconstruction are four spatial density distributions, one momentum
distribution and $\langle \hat a_0 \hat a_0 + \hat a_0^\dagger \hat
a_0^\dagger \rangle$.\label{fig:st3amomr}}
\end{figure}

In $State \, 1$ and $2$, the true value of the anomalous second
moments have all been equal to zero at $t = 0$. In contrast, for the
squeezed state, the absolute value of these elements are comparable
with the normal second moments. The absolute value of the true and
reconstructed second anomalous moments are displayed in
Fig.~\ref{fig:st3amomr}.

\section{\label{sek:disk}Discussion}
Initially, a short comment on the choice of observables might be in
order. If there were no constant background condensate the momentum
distribution could in principle be found from the spatial
distributions at late times. Having to deal with this background,
however, one has to include the additional observables in the MAXENT
density operator to exclude the possibility of the flat part of the
condensate being an almost even distribution of particles with all
allowed types of wave numbers whirling left and right, which would
otherwise be preferred by the MAXENT formalism.

Indeed we find for $State \, 1$ a very good reconstruction, the
errors being of the same magnitude as computer roundoff errors, from
using just measurements of the distribution in coordinate space at
three times and the number of particles in the zero wave number mode
$\langle \hat a^\dagger(0) \hat a(0) \rangle$. In addition to $State
\, 1$ in section \ref{sek:resultater}, we also tried to reconstruct
states of this type with more complicated initial distributions, but
still describable by a simple Gross-Pitaevskii wave function. These
included perturbations with two peaks and perturbations with one
peak and one hole in the flat background. In all cases the
reconstruction had the same precision as $State \, 1$, indicating
that the second moments, including the one-body density matrix, of
states describable by a simple Gross-Pitaevskii equation can be
completely reconstructed.

For $State \, 2$, we found that a wave number distribution also had
to be included as observable to get a somewhat reliable
reconstruction of the second moments. It is reasonable that it is
difficult to tell apart whether the moving particles belong to the
perturbation on top of a flat condensate or to some collective
motion of the whole, for example similar to $State \, 1$. Including
the momentum distribution remedied this problem somewhat, but errors
still persist (see Fig.~\ref{fig:st2nmom}). Subtracting the flat
background, these errors are of magnitudes about 10\% of the
perturbation peak value.

For $State \, 3$ with large numerical values of the anomalous second
moments, we find a reasonable reconstruction at $t = 0$ of the
normal second moments, but reconstruction of the anomalous second
moments completely fails, unless we additionally include the
observable $\hat a_0 \hat a_0 + \hat a_0^\dagger \hat a_0^\dagger$
in the MAXENT density operator, as we did in section
\ref{sek:resultater}.

A squeezed condensate may be useful in atom interferometry as input
to the dark side of an atomic mirror, whereupon a bright coherent
input will be split with number fluctuations below the values for a
binomial distribution \cite{uffeogklaus}. It is in a similar setup
that $\langle \hat a_0 \hat a_0 + \hat a_0^\dagger \hat a_0^\dagger
\rangle$ can be determined experimentally.

It is also worthy of notice that there was almost no improvement in
the reconstruction from including additional position and momentum
distributions for more points of time, and that using also the
observable $\langle \hat a_0 \hat a_0 + \hat a_0^\dagger \hat
a_0^\dagger \rangle$ for $State \, 1$ and $2$ did not change the
precision of the reconstructed second moments.

As for states significantly different from the ones treated here,
distributions in coordinate space at more points of time may be
required for a reliable reconstruction. Since enlarging the number
of measured distributions in space is rarely a problem, we suggest
for specific application to gradually include more distributions
until the results have stabilized. The calculation time is quite
manageable, the shown examples all taking well under an hour on an
ordinary PC.

\section{\label{sek:konklusion}Conclusion}
We have shown how to combine the MAXENT principle with the
Bogoliubov approximation to give reliable estimates of the normal
and anomalous second moments of the ladder operators, e.g. $\langle
\hat \psi^\dagger(x) \hat \psi(x') \rangle$ and $\langle \hat
\psi(x) \hat \psi(x') \rangle$, in a perturbed condensate. For
states describable by a simple Gross-Pitaevskii equation, we find
near perfect reconstruction from data of density distributions at a
few points of time and the number of particles in the zero wave
number mode. For states with a larger amount of particles in
non-zero wave number modes compared to the size of the perturbation
in coordinate space, we find it necessary to include a momentum
distribution to obtain a somewhat reliable reconstruction. Errors in
this case are less than 10\% of the perturbation peak value (peak
above the constant condensate background). Thus, the method makes it
possible to distinguish between a thermal and a coherent
perturbation of the condensate. Finally, we find that by measurement
of a single observable related to squeezing, apart from measurements
of momentum and coordinate space distributions, the method is able
to correctly reconstruct all second moments of the ladder operators
for squeezed states. \linebreak

\begin{acknowledgments}
We are grateful to Thomas Krogh Haahr Lynderup for technical advice
and to professor Hans C. Fogedby for useful discussions.
\end{acknowledgments}

\appendix
\section{Computational details \label{app:tipsogtricks}}
Several numerical problems may arise in finding the set $\{
\lambda_\nu \}$ that satisfies Eqs.~(\ref{eqs:cond}). As already
touched upon towards the end of section \ref{maxentteori}, we have
treated the problem as a minimization of deviations. We have chosen
to minimize the vector of deviations Eq.~(\ref{eq:fejlvektor}) by
using the Levenberg-Marquardt algorithm in the Matlab package. The
most common problem encountered when using this approach is that for
a given trial set $\{ \lambda_\nu \}$, the matrix $\mathscr P$ in
equation Eq.~(\ref{eq:rhomemin}) may not be positive definite. This
problem can be remedied somewhat by testing for positiveness using a
Cholesky factorization, which takes negligible time compared with
the para-unitary diagonalization. A better solution is to make a
good initial guess to the values of $\{ \lambda_\nu \}$. One can do
this by first running the minimization with an almost uniform
momentum distribution (in case this is included as observables) or
with $\langle \hat a_0^\dagger \hat a_0 \rangle$ set equal to
$N_{tot}/N$, and then carrying out the subsequent program runs using
the previously obtained $\{ \lambda_\nu \}$ as initial values. In
the subsequent program runs, one then gradually lets the momentum
distributions approach the real distribution. A similar approach may
be taken with the total number of particles, as making good guesses
of the $\{\lambda_\nu\}$ may be easier at low particle numbers.

A trick we have made use of, which made guessing the
$\{\lambda_\nu\}$ easier, is scaling down of the measured data.
Since we are not interested in the true many-body density operator
for the system, but only the second moments of the ladder operators,
we may as well make the calculations for a smaller number of
particles, provided the calculations are still precise, and then
rescale the data afterwards. As an example, we shall show how to
perform this down scaling in the b-basis at $t = 0$.

Let $\langle \bm b \bm b^\dagger \rangle$ be the matrix of (unknown)
second moments we wish to reconstruct. We will instead perform
calculations on the matrix $\langle \bm b \bm b^\dagger
\rangle_{DS}$, also having the characteristics of a matrix of second
moments:
\begin{eqnarray}\label{eq:downscale}
\langle \bm b \bm b^\dagger \rangle_{DS} &=& \kappa \left(\langle
\bm b \bm b^\dagger \rangle - \left(
\begin{array}{cc}
  I_N & 0 \\
  0 & 0 \\
\end{array}
\right)\right) + \left(
\begin{array}{cc}
  I_N & 0 \\
  0 & 0 \\
\end{array}
\right) \nonumber \\
&=& \kappa \langle \bm b \bm b^\dagger \rangle + (1-\kappa)\left(%
\begin{array}{cc}
  I_N & 0 \\
  0 & 0 \\
\end{array}%
\right)
\end{eqnarray}
where $I_N$ is the $N \times N$ identity matrix and $\kappa$ is a
real number. When reconstructing the matrix $\langle \bm b \bm
b^\dagger \rangle_{DS}$ we must modify the measurements, retaining
$gN_{tot}$ and thereby the transformation matrices $\mathscr A,
\mathscr B \text{ and }\mathscr U$. So, for instance, the down
scaled spatial density distributions that should be used can be
found from:
\begin{eqnarray}
\langle \bm \psi(t) \bm \psi^\dagger(t) \rangle_{DS} &=& \kappa
\langle \bm \psi (t) \bm \psi^\dagger (t)
\rangle + (1-\kappa)\cdot \nonumber \\
&& \mathscr A^{-1} \mathscr B^{-1} \left(
\begin{array}{cc}
  I_N & 0 \\
  0 & 0 \\
\end{array}%
\right) \left(\mathscr A^{-1} \mathscr B^{-1}\right)^\dagger.
\nonumber
\end{eqnarray}
Comparing the lower half part of the diagonals, we see that we
should multiply the coordinate space density measurements by
$\kappa$ and add to them $\frac{(1-\kappa)}{N}\sum_q v_q^2$. In a
similar manner, measurements of other observables should be modified
in the scaled down calculations. After the ${\lambda_\nu}$ has been
found, the relation Eq.~(\ref{eq:downscale}) can easily be inverted,
using the same point of time and basis as used in the original
scaling.

\section{Wigner functions \label{app:wignersjov}}
It is common practise to map the $N\times N$ one-body density matrix
to the real Wigner function $W(n,q)$ given on an $N\times N$ grid
\cite{Leonhardt}.
\begin{eqnarray}
W(n,q) &=& \frac{1}{N}\sum^{M}_{y =
-M}\rho\left(f(n-y),f(n+y)\right)\cdot \nonumber \\
& & \qquad \qquad \exp{\left(4\pi iyn/N\right)} \\
\nonumber \\
f(n-y) &=& \mod(n-y,N)-M-1 . \nonumber
\end{eqnarray}
The two arguments $n$ and $q$ can both assume N values, in our case
$\{-M,-M+1,\ldots,M\}$, like in Eq.~(\ref{eq:nvaerdier}). The
mapping is bijective so the Wigner function contains all the
information in the one-body density matrix. Furthermore, the wave
number- and spatial densities are easily recoverable:
\begin{eqnarray}
\langle\hat\psi^\dagger(x_n)\hat\psi(x_n) \rangle &=& \sum_{q=-M}^{M}W(n,q) \nonumber \\
\langle\hat a^\dagger(k_q)\hat a(k_q) \rangle &=&
\sum_{n=-M}^{M}W(n,q) . \nonumber
\end{eqnarray}
In the case of only one particle, the evolution of the quantum
system is completely described by the Hamiltonian and the one-body
density matrix. In our case, having many particles, we instead have
to specify all second moments to know the evolution of these, see
Eq.~(\ref{eq:startbet}). Therefore, in this paper, the Wigner
function is merely meant as an illustration of the one-body density
matrix at a particular point of time.

\end{document}